# Quantifying the Morphologies and Dynamical Evolution of Galaxy Clusters. II. Application to a Sample of ROSAT Clusters

David A. Buote

Massachusetts Institute of Technology

John C. Tsai

Canadian Institute for Theoretical Astrophysics

## ABSTRACT

We quantify the morphologies and dynamical states of 59 galaxy clusters using the power-ratio technique of Buote & Tsai applied to ROSAT PSPC X-ray images. The clusters exhibit a particularly strong $P_2/P_0 - P_4/P_0$ correlation in the $1h_{80}^{-1}$ Mpc aperture which may be interpreted as an evolutionary track; the location of a cluster on the correlation line indicates the dynamical state of the cluster and the distribution of clusters along this track measures the rate of formation and evolution of clusters in our sample. The power ratios anti-correlate with the cooling-flow rate indicating a reasonable dependence of the flow rate on cluster morphology. The relationship of the power ratios to the optical Bautz-Morgan (BM) Type is more complex. This is because the power ratios are sensitive to unrelaxed regions of clusters within a specified scale, whereas BM types are sensitive to unrelaxed regions over many scales. We discuss further astrophysical applications exploiting the relationship between the power ratios and the evolutionary states of clusters.

## 1. Introduction

The important connection between the morphologies of galaxy clusters and the cosmological density parameter $\Omega$ has received much recent attention (Richstone, Loeb, & Turner 1992; Evrard et al. 1993; Mohr et al. 1995). This connection has generally been formulated in terms of the frequency of "substructure" in clusters, where "substructure" is an ambiguous statement of the dynamical youth of a cluster (see West [1990] for a discussion of this issue). For example, Jones & Forman (1992) attempted to devise a more consistent description of "substructure" by visually separating clusters into six morphological classes. Using a sample of $\sim 200$ clusters observed with *Einstein*, Jones & Forman computed the frequency of these morphological classes and concluded that about 40% of clusters displayed some type of "substructure". Recent studies suggest a substantial increase in this percentage (e.g., West 1995).

From these qualitative measures of "frequency of substructure" in clusters, investigators have attempted to determine $\Omega$ (e.g., Richstone et al. 1992) and the power spectrum of primordial density fluctuations (e.g., David et al. 1993) by comparison to Press-Schechter (1974) type predictions of the distribution of collapsed (i.e. virialized) objects. However, these comparisons are inherently uncertain because of the unknown relationship between a particular investigator's definition of "substructure" and the dynamical state of a cluster. Even well defined and quantitative measures of morphology which employ centroid-shifts and axial ratios (Evrard et al. 1993, Mohr et al. 1995) do not clearly provide any physical connection of these parameters to the dynamical states of clusters.

We previously presented a method to quantify the morphologies of galaxy clusters in relation to their dynamical states as given by their gravitational potentials (Buote & Tsai 1995b; hereafter BT). The statistics of this method, i.e. "power ratios", in essence measure the square of the ratio of higher-order multipole moments of the two-dimensional potential to the monopole moment. The power-ratio description of the morphologies of X-ray clusters is intended to classify structure that is obvious to the eye, not subtle substructure that requires more robust techniques (e.g., Bird & Beers 1993); i.e. the significance of the substructure is a given, what the structure implies for the aggregate cluster dynamics is our concern.

Because of their intimate link to cluster dynamics, power ratios are not only ideal for cosmological studies, but also for studies of clusters themselves. It is our purpose in this paper to create a database of power ratios for testing the predictions of various cosmogonies and to present initial results on the implications of the measured power ratios. We discuss the cluster sample in §2 and the data reduction and analysis in §3. We consider correlations of the various power ratios among themselves and how these relate to Jones – Forman classes in §4. We present the evolutionary track of clusters and the correlations of the power ratios to other X-ray cluster properties and optical measures of substructure in §5. Further important astrophysical applications of the power ratios are discussed in §6, and we present our conclusions in §7.

## 2. The Sample

In principle one desires a large ($> 100$), complete, volume-limited sample of high signal-to-noise ($S/N$) X-ray images of clusters for use in cosmological analysis. Although not ideal, X-ray images of clusters taken with the ROSAT Position Sensitive Proportional Counter (PSPC; see Pfeffermann et al. 1987) are the best data currently available for a relatively large number of clusters (see BT §4 for a discussion). Ebeling (1993) has compiled a flux-limited sample ($\sim 200$ members) of nearby ($z < 0.2$) Abell and ACO clusters from PSPC data taken during the ROSAT All-Sky Survey (RASS); this sample, however, is estimated to be at most only 72% complete to a limiting flux $6.1 \times 10^{-12}$ erg cm$^{-2}$ s$^{-1}$. Unfortunately, the RASS is not available to the public and, in any event, the images of the RASS are not suitable for our needs because of the short exposures ($\sim 500$s). Long exposures of $\sim 50$ of the clusters in Ebeling's sample were taken as part of the Guest Observers (GO) program and are currently available in the ROSAT Public Data Archive operated by the HEASARC-Legacy database. However, only $\sim 30$ of the brightest clusters of Ebeling's sample are well represented by GO observations.

Since we have good coverage for only $\sim 30$ of the $\sim 200$ clusters in Ebeling's sample, we also consider the X-ray flux-limited sample of Edge et al. (1990). This sample of the brightest 55 clusters from EXOSAT and *Einstein* data is estimated to be $\sim 90\%$ complete and most are included in Ebeling's catalogue. We analyze all of the clusters in the ROSAT archive (subject to aperture-size constraints – see below) that are listed in the catalogues of Ebeling and Edge et al.. We also include the clusters A1413 and A2163 as part of the Edge et al. sample since David et al. (1993) showed that they were observed by *Einstein* to have fluxes above the Edge et al. flux limit. Given that our composite sample is not complete we will, in addition to presenting results for the whole sample, emphasize results for those clusters common to the (updated) Edge et al. sample.

Not all of the GO cluster images are useful for our analysis of the power ratios. Most importantly we require that a circle of at least a $500 h_{80}^{-1}$ kpc ($H_0 = 80 h_{80}$ km s$^{-1}$ Mpc$^{-1}$) radius be entirely enclosed within the central $40'$ diameter ring supporting the PSPC window. We only analyze regions interior to the PSPC ring because the support structure would contribute to the power ratios.

In Table 1 we list our sample of ROSAT PSPC clusters along with relevant data for each cluster. The fluxes listed in the table correspond to the $0.1 - 2.4$ keV band and have been taken from Ebeling where possible. For those clusters not in Ebeling's catalogue we compute a flux using the standard IRAF-PROS software; i.e. we extract the background-subtracted spectrum within the smaller of a radius of $1.5 h_{80}^{-1}$ Mpc or the interior to the PSPC ring and then fit an absorbed, single-temperature Raymond-Smith plasma having 1/2 solar metallicity. In addition to the flux, we give the redshift, the exposure time for the observation, background count-rate used in this paper, and the Bautz-Morgan Type (see §5.3). For those clusters with multiple exposures of the pointing we give the total exposure time of the merged observations. For Coma (A1656) and A2256 we refer the reader to the discussion of individual clusters in Appendix A.

## 3. Image Reduction and Analysis

To prepare the images for analysis we (1) eliminated time intervals of high background count rate, (2) selected energy channels corresponding to photon energies 0.5 - 2 keV and rebinned the image into $512 \times 512$ fields of $15''$ pixels, (3) corrected for exposure variations and telescopic vignetting, (4) merged multiple pointings (when available) into one image, (5) removed point sources (see §3.1), and (6) subtracted the background. All the data reduction was implemented with the standard IRAF-PROS software. In Appendix A we note the reduction peculiarities for each cluster.

The GO observations were partitioned into many short exposures to maximize the observing efficiency of ROSAT. We eliminated time intervals corresponding to large, short-term enhancements in the background light curves indicative of contamination from scattered X-rays, especially from the Sun, the bright Earth, and the SAA. Only a few clusters required time-filtering (Appendix A).

To minimize the effects of the X-ray background and the width of the PSPC PSF (see Hasinger et al. 1994) we selected photons only from energy channels between 0.5 and 2 keV. In addition, we rebinned the images into more manageable $15''$ pixels corresponding to $512 \times 512$ fields in accordance with BT. This pixel scale is the same as the exposure maps provided with the standard analysis systems software (SASS); note that the true resolution of the exposure maps actually corresponds to $30''$ pixels.

The images were then flattened using the SASS exposure maps. When dividing the images by these exposure maps we corrected for both exposure variations and telescopic vignetting. In principle this correction depends on the energy of each individual photon, but for energies above 0.2 keV the energy dependence is small and we neglect it (Snowden et. al. 1994).

For the few clusters having multiple pointings we merged them into one image for each cluster. Point sources common to each of the images for a particular cluster were used to align the fields. After removing point sources (see §3.1) we then subtracted the background. For most of the clusters we computed the background in a source-free region $\sim 45' - 50'$ from the field center.

### 3.1. Source Removal

Excising bright point sources from the cluster images is perhaps the most critical aspect of the image reduction. Because of the width of the PSPC PSF ($FWHM \sim 30''$) point sources are endowed with finite spatial extent and may contribute substantially to the power ratios (see §3.2) For the low-order power ratios that we consider the point sources do not appreciably affect the power ratios when the aperture radius $\gg$ width of the PSPC PSF. However, when this is not the case, precise source removal becomes vital.

As discussed in BT, removing only the brightest ($\gtrsim 5\sigma$) sources is generally sufficient for obtaining reliable measurements of the power ratios. We easily and effectively locate these brightest sources from visual examination of the cluster images. Nevertheless, in an effort to eliminate any non-cluster contamination, we also excise any dubious fluctuations not obviously associated with the host cluster. When done properly this only serves to smooth out the cluster image on small scales without altering the large-scale structure of the image (important to the power ratios under consideration – see §3.2).

From visual examination of the X-ray images it is impossible to unequivocally distinguish foreground/background sources from structures that are gravitationally associated with the clusters. However, the vast majority of the clusters in our sample have a single dominant component or two components that are sufficiently extended and bright which dominate the power ratios irrespective of our decision to include or exclude a few faint sources. Only for a few clusters (e.g., A500, A2382, and A514 – see Appendix A) having complex spatial structures is identification of real structures important. Even for these cases, though, the decision to include/exclude a source does not alter the power ratios to such a degree so as to give a complex cluster like A514 power ratios appropriate to a smooth cluster like A2029 (see §4). In any event, to achieve the most stringent constraints on the power ratios it is necessary to verify whether features in the complex clusters are indeed associated with the cluster; i.e. from detailed X-ray spectral analysis of the features and/or obtaining the appropriate redshifts at optical wavelengths.

We removed sources from the cluster images using the following simple procedure. For each source we constructed a circular annulus about the center of the source; typical annuli widths were 1-2 pixels. Then we fit a second-order polynomial surface to the background in the annulus. Finally, we replaced the source region with this background surface. In Figure 1 we show the image for A1795 before and after the sources have been removed. Of all the clusters in our sample A1795 has the largest number of detected sources within an aperture (i.e. $1h_{80}^{-1}$ Mpc) used to compute power ratios. The possibilities of biases due to excising sources from the images is discussed in §3.3.

### 3.2. Computation of Power Ratios

The power ratios are derived from the multipole expansion of the two-dimensional gravitational potential, $\Psi(R, \phi)$, due to matter interior to $R$,

$$\Psi(R, \phi) = -2Ga_0 \ln\left(\frac{1}{R}\right) - 2G \sum_{m=1}^{\infty} \frac{1}{mR^m} \left(a_m \cos m\phi + b_m \sin m\phi\right), \tag{1}$$

where the moments $a_m$ and $b_m$ are given by,

$$a_m(R) = \int_{R' \leq R} \Sigma(\vec{x}') (R')^m \cos m\phi' d^2 x',$$

$$b_m(R) = \int_{R' \leq R} \Sigma(\vec{x}') (R')^m \sin m\phi' d^2 x',$$

and $\vec{x}' = (R', \phi')$. For analysis of X-ray cluster images we associate the surface density, $\Sigma$, with the X-ray surface brightness; we refer the reader to BT for a more detailed discussion. By integrating the magnitude of each term of $\Psi$ over a circular aperture of radius $R_{ap}$ we arrive at the following definition,

$$P_m(R_{ap}) = \frac{1}{2\pi} \int_0^{2\pi} \Psi_m(R_{ap}, \phi) \Psi_m(R_{ap}, \phi) d\phi, \tag{2}$$

where $\Psi_m$ is the $m$th term in the expansion of eq. (1). The quantity $P_m$ is the "power" within $R_{ap}$ of the multipole terms of order $m$. Ignoring factors of $2G$, the powers are given by,

$$P_0 = [a_0 \ln(R_{ap})]^2, \tag{3}$$

for $m = 0$ and

$$P_m = \frac{1}{2m^2 R_{ap}^{2m}} \left(a_m^2 + b_m^2\right) \tag{4}$$

for $m > 0$. The total power of all multipole moments is simply,

$$P = \sum_{m=0}^{\infty} P_m. \tag{5}$$

The values for the individual $P_m$ depend on the coordinate system chosen. We utilize coordinate systems where the aperture is located at (1) the centroid of the cluster emission (i.e. where $P_1$ vanishes), and (2) at the peak of the cluster emission. To distinguish between these two cases we denote the moments of case (2) by $P_m^{(pk)}$. We consider case (2) in order to extract information from the dipole moment; i.e. $P_1^{(pk)}$ is akin to a centroid-shifting power, a quantity already known to be significant for many clusters (Mohr et al. 1993; Mohr et al. 1995).

We consider $P_m(m = 0, 1, 2, 3, 4)$ and $P_m^{(pk)}(m = 0, 1)$ for quantifying the morphologies of galaxy clusters. BT demonstrated that these powers are sensitive to the type of substructure most relevant to the dynamics of clusters and hence to cosmology (see Richstone et al. 1992). Rather than the powers individually, the power ratios, $P_m/P_0$ and $P_1^{(pk)}/P_0^{(pk)}$, classify clusters according to the dynamical importance of substructure (see BT). For example, consider a widely separated bimodal cluster of equal-sized (and spherical) components and a single-component ellipsoidal cluster having the same luminosity. Assume these clusters also happen to have identical values of $P_2$ within an aperture, $R_{ap}$. The bimodal cluster in this case necessarily has a smaller monopole moment within $R_{ap}$, and hence a smaller $P_0$, than the ellipsoidal cluster. Hence, $P_2/P_0$ differs for these clusters as a result of the different dynamical importance of substructure present in the clusters. Considering the power ratios $P_m/P_0$ and $P_1^{(pk)}/P_0^{(pk)}$ also has the advantage of normalizing to the cluster fluxes within $R_{ap}$, thus allowing consistent comparison between images of clusters having different fluxes and/or exposures.

We compute the power ratios in apertures of radii 0.5, 1, and 1.5 $h_{80}^{-1}$ Mpc under the condition that all points on the aperture boundary be separated by at least 1 pixel from the inside of the PSPC ring; for clusters just falling short of this criteria we decreased the appropriate aperture sizes up to 10% to allow computation of the power ratios. This criterion yields measurements of power ratios in the $0.5 h_{80}^{-1}$ Mpc aperture for all 59 clusters, 44 clusters have $1 h_{80}^{-1}$ Mpc aperture measurements, and only 28 have power ratios measured in a $1.5 h_{80}^{-1}$ Mpc aperture. By using apertures of different sizes we obtain information regarding the scale where the substructure is dynamically important; this is particularly useful for assessing where the gas is relaxed in a cluster (see §6).

### 3.3. Estimation of Uncertainty

We investigated the possibility of a bias introduced into the power ratios resulting from excising point sources from the cluster images. Returning to the case of A1795, we computed the power ratios in the $1h_{80}^{-1}$ Mpc aperture as a function of the the number of sources removed in decreasing order of brightness. The results are displayed in Figure 2. After the brightest few sources are removed the power ratios approach a stable solution, with the residual variations being substantially less than the uncertainties due to noise (see below and Table 2). Since A1795 encloses the most detected sources in our sample, any biases resulting from our source removal should be most pronounced. Hence, we conclude that the power ratios are not significantly biased as a result of our method for subtracting sources from the cluster images.

We applied a simple Monte Carlo procedure to estimate the uncertainties on the power ratios due to point sources and Poisson noise. Our starting point for a particular cluster was the reduced image except that the background was not yet subtracted. Following BT we added to the images point sources having spatial properties consistent with the PSPC PSF and numbers consistent with the $\log N(>S)$ - $\log S$ distribution given by Hasinger (1991). We excised the brightest simulated sources corresponding to (on average) the same number of point sources removed from the real image. Poisson noise was then added to the images. Since pixels having "0" counts represent poor estimates of the Poisson mean, we instead smoothed the image with a Gaussian ($\sigma = 15''$; i.e 1 pixel) which is essentially the width of the on-axis PSF. We find that the derived ranges of power ratios for the majority of clusters are not very sensitive to values of $\sigma \lesssim 2$ pixels. (We prefer to use the real image instead of an elliptical $\beta$ model [Mohr et al. 1995] since many of the clusters are not well described by the $\beta$ model.) After the sources and noise have been added we then subtract the mean background and compute the power ratios.

For each cluster we performed 100 realizations. We defined the 90% confidence limits on a given $P_m/P_0$ to be the 5th smallest and the 5th largest values obtained from the 100 simulations. Although this definition is arbitrary it serves our purpose for providing a simple, realistic measure of the significance of a given power ratio. In a few cases the derived confidence limits do not enclose the actual value of the power ratio. This is expected since on average 10% of the power ratios should not be enclosed by the 90% confidence limits. Moreover, our arbitrary definition of the confidence intervals may be inadequate in a few cases if the power ratios in the 100 simulations deviate substantially from a distribution symmetric about its mean.

### 4. Correlations of Power Ratios and The Jones – Forman Morphological Classes

We list the power ratios and their 90% confidence estimates in Table 2. The clusters in our sample generally span two decades in the various power ratios; i.e. $P_1^{(pk)}/P_0^{(pk)} \sim 10 - 1000$, $P_2/P_0 \sim 1 - 100$, $P_3/P_0 \sim 0.05 - 5$, and $P_4/P_0 \sim 0.01 - 1$ (all power ratios given in units of $10^{-7}$). Typically $P_1^{(pk)}/P_0^{(pk)}$, $P_3/P_0$, and $P_4/P_0$ have half-decade uncertainties, although $P_4/P_0$ appears to be slightly better constrained than $P_3/P_0$ on average. The best determined power ratio is $P_2/P_0$ which is generally constrained to better than a tenth of a decade; i.e. the even power ratios are more precisely measured than the odd ones.

The values listed in Table 2 can be used to construct individual distribution functions for the four power ratios considered here. Specifically, for each $P_m/P_0$ where $m = 1, ..., 4$, the distribution of the number of clusters which have power ratios of given values can be determined. These distribution functions can then be used to constrain differing aspects of the morphology of clusters produced by cosmogonic models, much as the observed luminosity function of galaxies is currently used to constrain theories of galaxy formation. This manner of comparison to theory will be investigated in a subsequent paper. Significant insights into cluster evolution and structure, however, can be more readily gained by considering correlations of the various power ratios.

Consider the three dimensional space defined by the centroided power ratios $(P_2/P_0, P_3/P_0, P_4/P_0)$. For the moment we exclude $P_1^{(pk)}/P_0^{(pk)}$ because it is necessarily trivially correlated with some combination of the $P_m/P_0$. Projections of the power ratios of the "best measured" clusters in Table 2 onto the three coordinate planes of power ratio space are shown in Figures 3 and 4, where assumed aperture sizes are $0.5 h_{80}^{-1}$ Mpc and $1 h_{80}^{-1}$ Mpc, respectively. We do not consider the $1.5 h_{80}^{-1}$ Mpc aperture because there are unsatisfactorily few clusters that meet our precision criteria. Our criterion for selecting the "best measured" values is that $P_3/P_0$ and $P_4/P_0$ do not have error estimates that span larger than a decade unless their upper limit is $\leq 0.25 \times 10^{-7}$. This arbitrary upper limit is used because clusters with small values of $P_3/P_0$ and $P_4/P_0$ often have large fractional uncertainties but still occupy a region of power ratio space well separated from that occupied by the other clusters. Since $P_2/P_0$ is the most precisely determined power ratio, the plots do not discriminate whether a cluster is "best measured" based on the values of this ratio. Next to each correlation plot we give an equivalent plot for the subset of clusters also included in the Edge et al. (1990) sample.

Despite the small number of clusters having well measured $P_3/P_0$ or $P_4/P_0$ (i.e., $\sim 30$ and $\sim 25$ clusters in the 0.5 and $1 h_{80}^{-1}$ Mpc apertures, respectively), the power ratios are obviously correlated with each other. The most pronounced of these is the $P_2/P_0$ - $P_4/P_0$ correlation, especially for the $1 h_{80}^{-1}$ Mpc aperture. These correlations are highly significant as determined by applying the Spearman Rank-Order Correlation test (e.g., Press et al. 1989, §13.8) which gives probabilities of $\sim 10^{-10}, 10^{-7}$ for the $P_2/P_0$ - $P_4/P_0$ correlation (where a probability of "1" indicates that the quantities are uncorrelated) for respectively the whole sample and the clusters common to the Edge et al. (1990) sample. In fact, the $1 h_{80}^{-1}$ Mpc correlation is consistent with $P_2/P_0 \propto P_4/P_0$ to within the estimated uncertainties (except A665, see below). Although only the clusters having the most precise measurements are displayed, all of the clusters are consistent with the trends shown in Figures 3 and 4.

Since clusters of all observed morphologies are included in our sample (see below), a first result is that the above correlations provide structural constraints on individual clusters. For example, consider the $P_2/P_0$ - $P_4/P_0$ correlation. All observed clusters regardless of morphology (again except for A665) have roughly the same amount of quadrupole structure ($P_2/P_0$) as octupole structure ($P_4/P_0$), modulo some constant factor; note this correlation is not evident when computing the power ratios on simulated blank images with Poisson noise. Although it would be reasonable for complex clusters having significant quadrupole structure to also have a large amount of octupole structure, the observed correlation implies that these differently scaled structures must increase or decrease in direct proportion to each other. An imaginary cluster assembled and evolving by arbitrary means certainly need not satisfy these constraints as shown in Figures 6 and 7 of BT. In fact, the toy clusters in BT which lie far to the left of the correlation track represent bimodal models where the small component lies along the short axis of the dominant component which would not be expected if the cluster initially collapsed along its shortest axis (e.g., Lin, Mestel, & Shu 1965). Our correlations readily show that since only one cluster lies off the correlation line, any theory of cluster formation must produce a large majority of clusters which obey the correlation.

We can further understand these correlations by considering the location of specific clusters in Figures 3 and 4 and how the power ratios relate to the qualitative classification system of Jones & Forman (1992; hereafter JF). We selected six clusters from our sample that span the complete range of power ratios and the six morphological classes of JF: A2029 – SINGLE, A1750 – DOUBLE, A85 – PRIMARY WITH SMALL SECONDARY, A2142 – OFFSET CENTER, A545 – ELLIPTICAL, A514 – COMPLEX. The dynamical states of these clusters is obvious: A2029 is a smooth, relaxed cluster; A1750 is a double cluster in the midst of a merger event; A85 has a dominant, mostly relaxed component and a small component about $\sim 600 h_{80}^{-1}$ kpc away that contains only a few percent of the total flux; A2142 possesses a center offset ($\sim 2'$) obvious from visual examination yet it only has a modest value of $P_1^{(pk)}/P_0^{(pk)}$; it is, however, quite elongated ($\epsilon \sim 0.35$ at a semi-major axis of $\sim 700 h_{80}^{-1}$ kpc) which clearly signals an unrelaxed state over large scales of the cluster (see, e.g., Buote & Tsai 1995a); A545 is the only cluster in our sample that is highly elongated but does not display any obvious center offset; A514 is highly irregular with several resolved mass components and is clearly in the very earliest stages of formation. We include A514, which is not in our original sample (see §2), because the JF COMPLEX class, of which A514 is the archetype, is not well represented in our sample. A2382, which appears to be the most COMPLEX cluster in our sample, is substantially more centrally condensed and smoother than A514. It will be useful to also consider the cluster A2319 (OFFSET CENTER) as a contrast to A2142 since the center displacement for A2319 translates to a substantial value of $P_1^{(pk)}/P_0^{(pk)}$ but the scale of the offset is confined to the core (i.e $\lesssim 300 h_{80}^{-1}$ kpc); i.e. the scale of unrelaxation differs for A2142 and A2319. We refer to the six clusters (A2029,A1750,A85,A2142,A545,A514) as the "reference" clusters and (A2319,A2382) as the "intermediate" clusters.

The JF classifications separate clusters based specifically on morphology and not on the scale over which that morphology exists. Obviously, the given morphological characteristics must exist within the aperture of the given observation by *Einstein* so there is some scale dependence present, but it is not evident what this scale is since observations of various clusters had different useful apertures. The power ratios specifically address the length scales over which the morphology is quantified by setting a consistent aperture size. We find that the power ratios evaluated on apertures of $0.5h_{80}^{-1}$ Mpc separate our sample of clusters along the lines of the JF classification scheme, hence the JF scheme qualitatively describes cluster morphology on $\sim 0.5h_{80}^{-1}$ Mpc scale. This is shown in Figure 5 where we plot the correlations of the $P_m/P_0$ computed in the $0.5h_{80}^{-1}$ Mpc aperture for the subset of clusters selected in the preceding paragraph.

The reference clusters are separated into two groups in the $P_2/P_0 - P_4/P_0$ plane. The group (A2029,A1750,A85) essentially appears smooth on this scale, since within the $0.5h_{80}^{-1}$ Mpc aperture only one of the subclusters is enclosed for A1750 and only the primary is enclosed for A85. However, by appealing to $P_3/P_0$ (see $P_3/P_0 - P_2/P_0$) these cluster are distinguished easily according to their dynamics; i.e. A1750 has the largest $P_3/P_0$, A85 an intermediate value, and A2029 the smallest $P_3/P_0$ as expected. The other group (A514,A545,A2142) appears unrelaxed in the $P_2/P_0 - P_4/P_0$ plane on this scale (i.e. guilt by association with A514). Again appealing to $P_3/P_0$, we see that A514 is well separated from the other two (A545,A2142) which themselves appear to have similar structure on the $0.5h_{80}^{-1}$ Mpc scale; this similarity is consistent with A545 ($z = 0.1540$) being similar in structure to A2142 ($z = 0.0899$) but less resolved. Note that the intermediate clusters A2382 and A2319 lie between these two groups but nearer the unrelaxed group as expected. For the reference and intermediate clusters $P_4/P_0 \propto P_2/P_0$ within the relatively large uncertainties for $P_4/P_0$ and thus, for the purposes of this discussion, do not appear to add significant information to that provided by $P_2/P_0$ alone. This is in contrast to $P_3/P_0$ which, when correlated with $P_2/P_0$, easily distinguishes the clusters according to their different stages of evolution within the $0.5h_{80}^{-1}$ Mpc aperture.

The power ratios computed in a $1h_{80}^{-1}$ Mpc aperture for the above subset of clusters do not in general classify clusters in accordance with the JF classifications, consistent with the above result that JF classes characterize cluster morphologies on a scale of $\sim 0.5h_{80}^{-1}$ Mpc. In Figure 6 we display the $P_2/P_0 - P_4/P_0$ correlation in the $1h_{80}^{-1}$ Mpc aperture for the reference clusters. The locations of some clusters have changed based on the structures that are most significant on the new length scale. For example, A1750 is now moved to the upper right hand corner because both subclumps of the cluster are now within the aperture. The breakdown of the JF classification system on this scale is easily seen by considering A2142 and A2319. These are both classified as OFFSET CENTER clusters by the JF scheme, however, they are well separated on the $P_2/P_0 - P_4/P_0$ correlation line indicating significantly different morphologies. In addition, the two clusters between A2142 and A2319 on the correlation line, A545 (ELLIPTICAL) and A85 (PRIMARY WITH SMALL SECONDARY), are of different JF classes.

## 5. Cluster Evolution and Correlations

### 5.1. The Evolutionary Track

The tight $P_2/P_0 - P_4/P_0$ correlation in Figure 6 as well as the placement of clusters of various morphologies on the correlation line allows a very simple interpretation for the plot. In the bottom–up cluster formation scenario, clusters in their infancy, born either via a merger event or the virialization of a single clump of material, have significant substructure and appear in the upper right hand part of the $P_2/P_0 - P_4/P_0$ plot. In the case of a merger, the cluster could be born as a double cluster like A1750. The non–axisymmetric structure gradually is erased as the cluster virializes, becoming a cluster like A2142, and then like A545. Finally the cluster becomes a single, well relaxed cluster such as A2029. The correlation line of Figure 6 (and Figure 4) is then interpreted as the track followed by clusters as they evolve from infancy to virialized states. Starting from the upper right, clusters move along the track down towards the lower left as they virialize. Alternatively, a cluster born as a single virializing object may first appear like A514 as a series of small substructures. These structures gradually agglomerate into a state represented by A2382. Possible continued accretion of small groups may lead to a mostly regular primary with a final renegade subgroup such as in A85. Finally, the subgroup merges and forms a completely relaxed cluster inhabiting the lower left part of the $P_2/P_0 - P_4/P_0$ correlation line.

Of course, evolution along the $P_2/P_0 - P_4/P_0$ sequence may proceed towards the upper right as well. A relaxed cluster like A2029 could subsequently accrete a small neighboring subcluster at which point it will be bumped back up to a position near A85 on the correlation line. Or, if a merger with a major secondary occurs, the merging cluster will occupy a position near that of A1750 (e.g., this may happen to A399 and A401 in $\sim 10^9$ yr).

In principle, it should be possible to distinguish the evolutionary tracks of the merging clusters (the double sequence) from that of the virializing single clumps (the complex sequence). Because the members of the complex sequence have more small scale power, we expect that for a given $P_2/P_0$, members of the complex sequence should have larger $P_4/P_0$ than members of the double sequence. The determinations of the power ratios, however, are not sufficiently accurate with currently available data to allow this distinction using only the $1 h_{80}^{-1}$ aperture. Perhaps the most straightforward approach to distinguish the evolutionary tracks is to combine information from apertures of different sizes; e.g., A514 and A1750 have very difference power ratios in the $0.5 h_{80}^{-1}$ aperture. Finally, note that the $P_2/P_0 - P_3/P_0$ correlation (Figure 6) could operate in the same manner as the $P_2/P_0 - P_4/P_0$ sequence if measured more accurately. The correlations for the power ratios in the $1.5 h_{80}^{-1}$ Mpc aperture for the "reference" clusters sufficiently distant to fit inside the PSPC central ring are entirely consistent with the evolutionary picture described for the $1 h_{80}^{-1}$ Mpc aperture, but with much larger uncertainties.

We comment on the one cluster in our sample that significantly deviates from the evolutionary picture described above. The outlier A665 (given by the point which lies farthest from the $P_2/P_0 - P_4/P_0$ correlation line of Figure 4), has large values of $P_1^{(pk)}/P_0^{(pk)}$ and $P_3/P_0$, and modest (but uncertain) values of $P_4/P_0$ typical of a cluster in our sample being dynamically unrelaxed on both $0.5h_{80}^{-1}$ Mpc and $1h_{80}^{-1}$ Mpc scales. However, the $P_2/P_0$ values are anomalously low and indicative of a very relaxed cluster on these scales. These characteristics are consistent with a state in which the X–ray emitting gas neither traces the total mass nor the the potential of the cluster. For example, it was shown (Buote & Tsai 1995a) that during the late time (i.e. $z = 0.83 - 0.13$) evolution of the Katz & White (1993) simulation, the simulated cluster experienced a brief period during a merger (i.e. $z \sim 0.5 - 0.3$) where the X-rays did not follow the dark matter distribution or the potential. During this time the X-ray isophotes were very distorted yet the ellipticity (estimated using quadrupole moments) was very small ($\epsilon \sim 0.15$), much less than at earlier ($\epsilon \sim 0.5$) and later ($\epsilon \sim 0.3$) times.

If this description for A665 is indeed correct, this cluster has been captured during a very interesting phase of cluster formation. The gas is undergoing the greatest dissipation in going from a distribution like that of the dissipationless component of the cluster (or being distributed in virial equilibrium with individual subclumps, such as for A1750) to that of following the total cluster potential. Clusters which lie higher than A665 on the evolutionary track must then have gas that is not following the total cluster potential and clusters lying below A665 must have gas virialized with the total potential. Since only 1 cluster in our sample of 59 clusters deviates from the evolutionary track, this epoch of cluster formation must be very short, as already hinted at by simulations (Buote & Tsai 1995a). The consideration of A665 also slightly modifies the picture of cluster evolution described above. That is, as clusters virialize and move toward the lower left on the correlation line, they experience a brief episode where they evolve off the track, and then fall back to it after the cluster gas has dissipated sufficiently to follow the overall potential.

### 5.2. Correlations with other X-ray Properties

We investigated correlations of the power ratios with X-ray temperature, X-ray luminosity ($0.1 - 2.4$ keV), and cooling-flow rate using data from the literature (see Table 1; Edge et al. 1990,1992; David et al. 1993; Fabian 1994). Although correlations with X-ray temperature and X-ray luminosity cannot be ruled out because of the generally small number of data points (i.e. $\lesssim 20$ for each quantity), there are no obvious correlations of these quantities with the power ratios. This lack of strong correlations is reasonable since the power ratios measure evolution without regard for the mass of the cluster to which these other quantities are sensitive (Edge & Stewart 1991), although we might expect a correlation if clusters of different masses evolve at different rates (e.g., Richstone et al. 1992). In accordance with this reasoning we find no clear evidence for a correlation of the power ratios with the optical velocity dispersions from Struble & Rood (1991).

Using the mass-flow rates from Fabian (1994) for the Edge et al. (1990;1992) clusters we find that the power ratios computed in both the $0.5h_{80}^{-1}$ Mpc and $1h_{80}^{-1}$ Mpc apertures are clearly anti-correlated with mass-flow rate; the weakest trend is observed for $P_1^{(pk)}/P_0^{(pk)}$. In Figure 7 we show the mass-flow rate vs $P_2/P_0$ computed in the $1h_{80}^{-1}$ Mpc aperture which exhibits the strongest correlation (Spearman Rank-Order Correlation probability of 0.08 – see §4); clusters with no detected mass-flow rates were placed on the bottom of the plot to show the range in $P_2/P_0$ for these clusters. A negative correlation of mass-flow rate with evolution (as given by $P_2/P_0$) is clearly reasonable if the rate of central cooling of gas can be reduced by recent mergers or by only recently virializing structure. Much of the scatter in Figure 7 may be due either to the large uncertainty in the mass-flow rates, which are accurate perhaps only to a factor of a few, or to a definite (albeit weak) correlation with X-ray luminosity (Edge et al. 1992). Nevertheless, the power ratios and central cooling rates give consistent pictures of the evolutionary states of galaxy clusters.

### 5.3. Comparison to Bautz-Morgan Type

We investigate whether the power ratios yield a description of the evolutionary state of the clusters consistent with the optical Bautz-Morgan (1970; BM) classification scheme. The BM type of a cluster measures, "the degree to which the brightest member stands out against the general cluster background"(BM). Although assigning BM types to clusters is inherently subjective and prone to systematic errors (see, e.g., Leir & van den Bergh 1977) the BM scheme generally provides some measure of the evolutionary state (e.g., Sandage & Hardy 1973; Leir & van den Bergh 1977). In Table 1 we give the BM types for clusters available from the literature. The BM types are uncertain to at least a half-type and those with a ":" are even less certain.

Since the BM scheme measures the cluster evolutionary state (Type I being the most relaxed, Type III most unrelaxed), we expect a positive correlation of the power ratios with BM type if both X–ray and galaxy distributions do indeed trace the evolution of the cluster. The BM scheme, however, does not specifically address different scales for a given cluster and only specifies the "dominance of the brightest members" globally. Furthermore, the BM scheme is unlike the JF classifications because this latter method does sort clusters by their morphology on scales $\sim 0.5h_{80}^{-1}$ Mpc (see §4), although this may not necessarily have been originally intended. To illustrate this difference we again consider A2319. Recall that the power ratios computed in the $0.5h_{80}^{-1}$ Mpc aperture classify A2319 as an unrelaxed cluster due to the substructure in the core, but in the $1h_{80}^{-1}$ Mpc aperture the gravitational effects of the core subclustering are unimportant and A2319 appears relaxed. The BM scheme in this case identifies the unrelaxed nature of the core of A2319 and classifies it as Type II-III. In contrast, A1750, which is classified by power ratios as relatively relaxed on small scales ($0.5h_{80}^{-1}$ Mpc) but unrelaxed at a scale of $1h_{80}^{-1}$ Mpc also has a BM type of II – III. We therefore expect the power ratio – BM correlations to be the superposition of a positive correlation for those clusters which have substructure on the scale currently specified by the power ratios and a negative correlation for those clusters which do not have substructure on the current scale, but do have substructure on a different scale.

We first consider the BM Types for the reference clusters examined in §4. The power ratios

$P_m/P_0$ computed in the $0.5h_{80}^{-1}$ Mpc aperture (Figure 5) clearly separate BM Type in the positive sense – larger power ratios imply larger BM Type. Although A545 (III) and A2142 (II), which are classified into essentially the same region of power-ratio space, differ in BM Type by 1, the disagreement is not highly significant considering the BM uncertainties. In the $1h_{80}^{-1}$ Mpc aperture (Figure 6), BM Type and the power ratios correlate extremely well with the exception of A545; i.e. A1750 and A514 (both II-III) are at the top of correlation line, A2142 (II) is in the middle, A85 and A2029 (both I) are at the bottom. However, since A545 is a good candidate for core substructure the different BM classification may be result of the scale-dependence effect we discussed above for A2319 and A1750. Overall the power ratios for our reference clusters and the BM Types correlate well.

The BM Type – power ratio correlation for the whole sample does not present such a clean picture. We focus on the BM – $P_2/P_0$ correlation since $P_2/P_0$ is the most precisely measured power ratio and it has the most straightforward evolutionary interpretation (for the $1h_{80}^{-1}$ Mpc aperture). In Figure 8 we plot the results for all the clusters and for those included in the Edge et al. (1990) sample. Both plots, particularly for the Edge et al. subset, show the expected superposition of the positive and negative correlation discussed previously. That is, the most relaxed clusters inhabit the bottom left of the plots and for slightly larger values of $P_2/P_0$ the BM Type also increases slowly from I to II. For clusters with large-scale substructure (A514,A1750) this positive correlation continues to the largest values of $P_2/P_0$ and BM (almost). A noticeable outlier in this positive correlation is A3558 classified as BM Type I. There is obvious substructure in the PSPC image (see Appendix A) which, if indeed gravitationally associated with the cluster, implies A3558 should be considered dynamically young.

Along with this positive correlation, however, there is a negative correlation that is particularly evident in the top left of the plot of the Edge et al. clusters. These clusters appear relatively relaxed on the current scale but have departures from equilibrium on smaller scales. Appealing to the other power ratios (which are not as well constrained as $P_2/P_0$) does not add any additional clarification when combined with the already uncertain BM description. Thus, to within the uncertainties present, both galaxies and X–rays consistently trace the dynamical state of clusters. Although the BM classifications are qualitative, the results of this section suggest that much can be learned from future joint considerations of structure in the gas and galaxies.

## 6. Discussion

If galaxy clusters form hierarchically then a simple interpretation of the strong $P_2/P_0 - P_4/P_0$ correlation in Figure 6 is a track followed by clusters as they evolve from infancy to virialized states (§5.1); i.e. the power ratios distill from the morphologies of galaxy clusters a clear, quantitative measure of their evolutionary states. Other methods that attempt to use morphologies to probe cluster evolution suffer because they neglect to weigh the intrinsic scales of the morphological features in proportion to the cluster gravitational potential. We now discuss some potential astrophysical applications of the power-ratio connection to cluster evolution.

Perhaps the most important use for the power ratios is for comparison with cosmological N-body / hydrodynamic simulations. As we discussed in BT (§6) the power ratios are ideally suited to test the Morphology – Cosmology Connection (MCC; e.g., Evrard et al. 1993; Mohr et al. 1995). This MCC simply states that the observed structure of X-ray clusters is sensitive to the cosmological density parameter $\Omega$, an idea suggested by, e.g., Richstone et al. (1992). However, it is actually the current merger rate (or formation rate), *which describes the current dynamical states of clusters*, that is particularly sensitive to $\Omega$ (see Figure 1 of Richstone et al.). As we stressed above (and in BT) morphology alone does not clearly indicate the dynamical state of a cluster unless the intrinsic scales of the morphological features are weighed appropriately. The distribution of power ratios (particularly $P_2/P_0$ in the $1h_{80}^{-1}$ Mpc aperture), which are by construction related to the dynamical state, provide a quantitative measure of the formation rate of clusters in our sample and may allow for more precise constraints on $\Omega$ to be obtained than methods which quantify morphologies without regard to dynamics.

Whereas the distribution of clusters along the tight $P_2/P_0 - P_4/P_0$ correlation in the $1h_{80}^{-1}$ Mpc aperture should suffice for testing the MCC, the joint use of individual distributions of the power ratios, $P_m/P_0$, can provide a further test of the MCC and can also give constraints on the primordial power spectrum on scales near that of clusters. The discovery of the evolutionary track also provides vital constraints on cluster formation in general; our interpretations of both placement and movement along the track can be tested in detail by hydrodynamical simulations of cluster formation. The unique position in cluster evolution occupied by A665 can be tested also by comparing the distribution of projected mass, as can be determined by weak lensing (e.g., Kaiser & Squires 1993), to the gas distribution as quantified by the present work.

Since the power ratios provide a measure of the evolutionary states of clusters, they may be used to search for and summarily remove the effects of evolution in particular investigations of clusters. For example, if the relative proportions of galaxy morphological types within clusters is correlated with evolution (i.e. the Dressler [1980] Morphology - Density relation; also see Whitmore [1990]) they should correlate with the power ratios. Distance indicators like brightest cluster galaxies (BCGs) and the universality of the luminosity function (Schechter & Press 1976) may also be affected by the evolutionary states of clusters. For example, Sandage & Hardy (1973) use the observed correlation between BCG magnitude and BM Type to reduce the scatter in their Hubble Diagram. The power ratios, which give a similar but much more precise measure of the state of a cluster (see §5.3), should be useful in further reducing the scatter in the Hubble Diagram and remove systematic evolutionary effects associated with BCGs (e.g., Weir, Djorgovski, & Bruzual 1990); i.e. in addition to correlating with $P_2/P_0$ in the $1h_{80}^{-1}$ Mpc aperture, it may be useful to simultaneously correlate with the $0.5h_{80}^{-1}$ Mpc aperture results if BCG magnitude is especially sensitive to departures from equilibrium in the core of the cluster. Unfortunately there are too few published BCG magnitudes for the clusters in our sample to allow us to meaningfully investigate this issue at present.

- 16 -The power ratios could provide a natural means to "correct" X-ray mass estimates of clusters due to departures from hydrostatic equilibrium. Along with weak gravitational lensing (e.g., Kaiser & Squiers 1993), X-ray images of clusters are the most powerful means to measure cluster masses (see, e.g., Mushotzky 1995). It would be straightforward to compute the errors in cluster masses assuming hydrostatic equilibrium using N-body / hydrodynamic simulations (e.g., Tsai, Katz, & Bertschinger 1994) of clusters spanning the narrow range of power ratios in Figure 4. This "template" may then be used to correct hydrostatic mass estimates of real clusters. Moreover, the power ratios computed in different apertures indicate where in a given cluster hydrostatic equilibrium is a good approximation; e.g., the power ratios for A2319 (§4) show that it is substantially more relaxed in the $1h_{80}^{-1}$ Mpc aperture than in the $0.5h_{80}^{-1}$ Mpc aperture and thus hydrostatic analysis is better suited for that larger scale to reduce the nonequilibrium effects of subclustering on small scales – this is precisely the argument previously made by us (Buote & Tsai 1995a). Hence, the power ratios can both indicate where it is best to apply hydrostatic analysis in clusters and, if the errors are indeed correlated with evolution, correct for nonequilibrium effects.

Power ratios may be more suitable for probing the dynamics of clusters than traditional X-ray methods that assume the gas is in hydrostatic equilibrium. For example, Kaiser (1991) predicts morphological differences between the X-rays and mass in high-redshift clusters ($z \sim 0.5$) that may provide important constraints on the nature of the dark matter. The power ratios allow a natural comparison of the dynamical state of a cluster as indicated by the X-ray gas with that indicated by the mass distribution obtained from weak lensing (e.g., Kaiser & Squiers 1993; Smail et al. 1995).

## 7. Conclusion

We have used the power-ratio technique (BT) to quantify the X-ray morphologies of a sample of 59 galaxy clusters. The power ratios quantify substructure in a cluster in relation to its influence on the gravitational potential, or, equivalently, the dynamical state of the cluster. Hence, we have in effect a measure of the evolutionary states of the clusters.

Our sample consists of clusters belonging to the X-ray flux-limited samples of Edge et al. (1990) and Ebeling (1993) that have sufficiently high $S/N$ observations with the ROSAT PSPC. Only for the brightest $\sim 30$ clusters is our sample approximately complete ($\sim 60\%$) to the Edge et. al. flux limit of $1.7 \times 10^{-11}$ erg cm$^{-2}$ s$^{-2}$.

We computed the power ratios in a circular aperture of radius $0.5h_{80}^{-1}$ Mpc for all of the clusters in the sample. For sufficiently distant clusters we also computed power ratios in $1h_{80}^{-1}$ Mpc and $1.5h_{80}^{-1}$ Mpc apertures when the entire aperture fit within the $40'$ diameter ring of the PSPC. We estimated 90% confidence uncertainties on the power ratios using a Monte Carlo procedure.

The power ratios exhibit striking correlations, particularly $P_2/P_0 - P_4/P_0$ in the $1h_{80}^{-1}$ Mpc aperture. From consideration of "reference" clusters spanning the full range of power ratios and belonging to the six morphological classes of JF, we interpreted this $P_2/P_0 - P_4/P_0$ correlation as an evolutionary track for clusters where young clusters arrive at the top of the track (i.e. large values of $P_2/P_0$ and $P_4/P_0$) and evolve downwards to small values of $P_2/P_0$ and $P_4/P_0$. The relative distribution of clusters along this track provides a quantitative measure of the current formation rate of clusters in our sample – a quantity that has been shown to be sensitive to $\Omega$ (Richstone et al. 1992). We also find that the JF classes qualitatively distinguish clusters on scales of $\sim 0.5 h_{80}^{-1}$ Mpc, as indicated by the reference clusters being similarly classified by power ratios on this scale.

We find no evidence for correlations of the power ratios with X-ray temperature, X-ray luminosity, or velocity dispersion; this lack of strong correlations is reasonable because the power ratios are not directly sensitive to the masses of clusters since the small evolutionary dependences on these quantities are dwarfed by the differences associated with the individual cluster masses (we might also expect a correlation if clusters of different masses evolve at different rates – e.g., Richstone et al. 1992). A negative correlation of the power ratios with mass-flow rate is observed and is consistent with a reasonable expectation that the rate of gas cooling in the cores of clusters is related to the dynamical state of the cluster. We also find that the optical BM classifications for clusters are related to the power ratios in an understandable manner. That is, whereas power ratios are sensitive to some given scale, the BM classes do not differentiate between clusters that are unrelaxed over large scales ($\sim 1 h_{80}^{-1}$ Mpc) and those that are substantially unrelaxed only in the core ($< 0.5 h_{80}^{-1}$ Mpc). The relation of BM types to power-ratio values suggests that both the gas distribution and the galaxy distribution reflect similar evolutionary states for a cluster.

We describe several astrophysical applications that exploit the connection between the power ratios and evolutionary states of clusters. In particular, we discuss the suitability of the power ratios for (1) constraining $\Omega$ via the Morphology – Cosmology Connection, (2) correcting distance indicators like BCGs for the effects of cluster evolution, and (3) correcting mass estimates of clusters for departures from hydrostatic equilibrium. In principle, given adequate observations, power ratios can be used to address any problem where cluster evolution is an issue.

The insights into cluster evolution gleaned from our relatively small sample of clusters, as well as the potential for precise cosmological constraints provided by power ratios, highlight the need for a much larger sample of high-quality X-ray data of clusters of galaxies. With the prospects of the XMM for studying low-redshift clusters and AXAF for high-redshift clusters it will in principle become possible to realize the full potential of the power ratios to be a potent astrophysical tool.

We benefited from discussions with E. Bertschinger, J. Blakeslee, C. Canizares, E. Gaidos, and J. Tonry. We express our gratitude to M. Corcoran for assistance with the ROSAT archive and to D. Harris for his advice regarding merging PSPC images. We acknowledge use of the following astrophysical databases: ADS, HEASARC-Legacy, NED, and SIMBAD. DAB was supported by grants NAG5-1656, NAS8-38249 and NASGW-2681 (through subcontract SVSV2-62002 from the Smithsonian Astrophysical Observatory).

## A. Notes on Individual Clusters

Here we list details of the image reduction for individual clusters with particular emphasis on point sources.

A21: This is a bimodal cluster whose two components are separated by $\sim 3'$ and are easily resolved by the ROSAT PSPC image. Short-term enhancements of the light curve were removed from the original image (exposure: 9068s) to yield an effective 8680s exposure time. One faint source straddling the 0.5 Mpc aperture was removed. An additional bright source was removed from the 1.5 Mpc aperture.

A85: It is the quintessential PRIMARY WITH SMALL SECONDARY cluster in the JF classification scheme. The small secondary structure lies $\sim 10'$ to the S. Five faint sources were removed within the ring. They lie within the 1 Mpc aperture but not the 0.5 Mpc aperture. The large sub-clump to the South is not removed.

A119: This cluster has an interesting tail of emission to the North that suggests significant departures from equilibrium. Two sources were removed from the 0.5 Mpc aperture. The 1 Mpc aperture did not fit inside the ring so we reduced the size to 940 kpc which then did fit inside. Six more sources had to be removed from this larger aperture.

A400: This irregular cluster has a dominant peak and a few secondary peaks. The nearest peak is $2'$ to the East of the dominant peak is a good candidate for a subcluster. We had to decrease the aperture size to 460 kpc in order to meet the criteria of §3. Five bright and four faint sources were removed from the 460 kpc aperture.

A401: A very smooth single-component cluster. A399 may be seen outside the PSPC ring to the SW. In order to avoid contamination from A399 we estimated the background in a circle of $5'$ radius $40'$ to the NW. Two pointings (exposures: 7465s and 6797s) on the cluster center were merged into one image. Two faint sources and the cluster center were used to register the images to a reference coordinate frame. The required shift was consistent with no shift in the E-W direction but $1.8 \pm 0.5$ pixels N-S which is slightly larger than the expected pointing errors of the PSPC. No sources needed to be excised from the 0.5 and 1 Mpc apertures, but the two aforementioned faint sources were removed from the 1.5 Mpc aperture. Note that the emission from A399 which lies off-axis $\sim 30'$ to the SW begins to become significant within the ring for the 1.5 Mpc aperture.

A478: Two sources were removed from this smooth-looking cluster in the 0.5 Mpc aperture, four more from 1 Mpc, and five more from the 1.5 Mpc aperture.

A496: From this symmetrical single-component cluster we removed four faint sources within the ring.

A500: This very irregular cluster has at least three distinct emission peaks within 0.5 Mpc in addition to the dominant central peak. The emission is roughly centered on galaxies associated with the cluster. Of the peaks near the center we removed only the largest source $\sim 4'$ to the NW because it did not have any clear association with the diffuse emission; follow-up observations need to be performed to determine the nature of these sources. Four bright sources were removed from the 0.5 Mpc aperture. Four bright and 12 faint sources were removed from the 1 Mpc aperture.

A514: This is the archetype COMPLEX cluster in the JF system. There are at least three distinct peaks in the central $\sim 5'$ cluster continuum and then two other peaks $10'$ to the W. All of the peaks appear to be extended and/or are part of the cluster emission; however this needs to be verified. No sources were removed from the 0.5 Mpc, one faint source from the 1 Mpc, and three bright sources from the 1.5 Mpc apertures.

A545: This cluster is highly elongated within the 0.5 Mpc aperture which may reflect dynamical youth of the interior. One bright source was removed from the 1 Mpc aperture and an additional bright and one faint source were removed from the 1.5 Mpc aperture.

A586: Although the cluster appears to be mostly smooth this may be the result of the relatively low $S/N$ of the observation. Two faint sources were removed from the 1.5 Mpc aperture.

A644: This looks like a smooth single component cluster. Two faint sources were removed from the 0.5 Mpc aperture. One bright and four faint sources were removed from the 1 Mpc aperture. In order to fit inside the ring (see §3) the last aperture was taken to be 1.45 Mpc. Two faint sources as well as some extended emission near the ring to the East were removed from this 1.45 Mpc aperture.

A665: This irregular cluster has its emission peak clearly displaced from the centroids of the outer isophotes. Starting from the emission peak the isophotes fan out to the North. Because of the long exposure time there are many point sources easily detected in the field. No sources needed to be removed from the 0.5 Mpc aperture; one bright source was removed from the 1 Mpc aperture; five faint sources straddling the 1.5 Mpc aperture were removed. It is interesting that $P_1^{(pk)}/P_0^{(pk)}, P_3/P_0$, and $P_4/P_0$ classify A665 as unrelaxed but $P_2/P_0$ is typical of smooth, relaxed clusters (see §4).

A754: The center is clearly offset from the outer cluster emission. No sources were removed within the 0.5 Mpc aperture. Because the center of the cluster emission is displaced from the field center, only the 0.5 Mpc fits within the PSPC ring according to our criteria from §3.

A1068: No sources needed to be removed from the 0.5 Mpc aperture of this regular-looking cluster. Two faint sources were excised from the 1 Mpc aperture, and one additional faint source was removed from the 1.5 Mpc aperture.

A1361: This cluster appears reasonably smooth but the $S/N$ is relatively low for our sample. While no sources were removed from the 0.5 Mpc aperture, one bright and one faint source were removed from the 1 Mpc aperture. One more bright and three faint sources were removed from the 1.5 Mpc aperture.

A1413: No sources were removed from the 0.5 Mpc aperture of this mostly regular cluster, but two bright and one faint source were removed from the 1 Mpc aperture. Three more faint sources were removed from the 1.5 Mpc aperture.

A1651: The cluster appears to be mostly smooth and regular. One bright source on the 0.5 Mpc aperture, two bright and one faint sources within the 1 Mpc aperture, and another faint source in the 1.5 Mpc aperture, were removed.

A1656: There were four pointings (exposure times: 22183s, 21893s, 20691s, and 22427s) on the Coma field each at a different position. Unfortunately these pointings were not placed symmetrically around the cluster center to accommodate the circular apertures required for the power ratios. Nonetheless, we needed to merge these observations to obtain sufficient pixels to even get the 0.5 Mpc aperture covered. Each of these observations showed evidence of short-term enhancements in their light curves which we removed resulting in exposure times of 20032s, 21893s, 20691s, and 22427s respectively. Four bright point sources were used to register the images to the coordinate frame of the 20032s image and then they were added together. As usual we included only the regions interior to the PSPC ring for each observation. Because of the dominant emission from the cluster we only removed one source within the 0.5 Mpc aperture. There are other sources in the Coma image found by White, Briel, & Henry (1993), but they are faint compared to the cluster continuum. In any event their extent is much smaller than the aperture size and should only contribute appreciable to multipole components higher than we are considering. We had to decrease the aperture size to 0.45 Mpc to meet our criterion of §3.1.

A1689: The X-rays appear smooth and symmetrical in this cluster that also is known to have giant arcs due to gravitational lensing of distant galaxies. No sources needed to be removed from the 0.5 and 1 Mpc apertures. One bright source and four faint sources straddling the 1.5 Mpc aperture were removed.

A1750: This is a double cluster with its components separated by $\sim 10'$ (i.e. $\sim 1$ Mpc). Because the centroid of the two nearly equal-sized subclumps moves with increasing aperture size the 1.5 Mpc aperture just touches the ring. We decrease the aperture to 1.4 Mpc to fit inside the ring within the specifications stated in §3; the power ratios are essentially unaffected by this reduction in aperture size. One source was removed from the 0.5 Mpc aperture, two more from the 1 Mpc aperture, and an additional 7 from the 1.5 Mpc aperture.

A1795: A perfect example of a smooth, single-component, regular-looking cluster. The image of this cluster within the ring is literally peppered with point sources. In all 20 sources were removed within the 1 Mpc aperture. Only two of the sources lie within the 0.5 Mpc aperture. The power ratios are not affected (within the Monte Carlo error estimates) when only the brightest few sources are removed.

A1837: The emission is highly elongated in the inner $5'$ of this cluster, but only a single peak is evident to the eye. Two bright and three faint sources were removed from the 0.5 Mpc aperture.

A1914: This cluster looks mostly regular and smooth. No sources needed to be excised from the 0.5 Mpc aperture, but three faint sources and an additional faint source were removed from the 1 and 1.5 Mpc apertures respectively.

A1991: Three faint sources were removed from the 0.5 Mpc aperture of this regular-looking cluster and three bright and five additional faint sources were removed from the 1 Mpc aperture.

A2029: This is a smooth, regular cluster. One source was removed from the 0.5 aperture, four more from the 1 Mpc aperture, and three additional sources from the 1.5 Mpc aperture.

A2034: The centroids of the X-ray isophotes appear to shift to the S at large radii of this somewhat irregular cluster. No sources were removed from the 0.5 Mpc aperture. One faint source was removed from the 1 Mpc aperture and 1 bright and an additional faint source were removed from the 1.5 Mpc aperture.

A2052: Two observations (exposure times: 6215s and 3032s) pointed on the cluster center were merged. One point source and the cluster center (which is very centrally peaked) were used to register the images to the coordinate frame of the 6215s image; the required shift was 1 pixel which is consistent with the pointing accuracy of ROSAT. Only the one source needed to be removed.

A2063: Two bright and four faint sources were removed from the 0.5 Mpc aperture of this mostly symmetrical and smooth cluster.

A2107: Mostly regular in appearance, we removed six faint sources from the 0.5 Mpc aperture.

A2199: This is another beautiful, regular, single-component cluster. Seven faint sources were removed from the 0.5 Mpc aperture.

A2142: This cluster is highly elongated and has an obvious center displacement from the outer emission; i.e. a classic OFFSET CENTER in the language of JF. We merged three observations (exposure times: 7740s, 6192s, and 4941s) that were pointed on the same coordinates. Each image was flattened first and then three point sources (one bright, two faint) were used to register the images to the coordinate frame of the 7740s image. The required shifts were less than one pixel. Once the images had been merged several more point sources become easily apparent. In particular, the bright source used for registration is 15 pixels NE of the cluster center and must be carefully excised from the cluster continuum. It is the only source removed from the 0.5 Mpc aperture. Five more were removed from the 1 Mpc aperture and another 8 from the 1.5 Mpc aperture.

A2163: This cluster, the most distant in our sample, was observed twice with the PSPC: wp800188 (5034s) and wp800385 (7099s). The first exposure required substantial time-filtering which left only 3426s of useful of exposure for wp800188. Three sources were used to align the fields which required no significant shift in E-W direction, but a $-1.4 \pm 0.51$ pixel shift in the N-S direction which is consistent with the pointing accuracy of ROSAT. The centroid of A2163 in the core appears to be offset from the outer isophotes. There are no sources within $1.0h_{80}^{-1}$ Mpc of the cluster. However, there are two sources about $1.5h_{80}^{-1}$ Mpc to the N of the cluster center which we removed.

A2204: No sources needed to be removed from this apparently smooth cluster.

A2218: This cluster is quite elongated but smooth in appearance. Because of the long exposure of the observation there are many resolved point sources in the field. However, because A2218 is at relatively large redshift, the 1.5 Mpc aperture only encloses a small portion of these sources. No sources needed to be removed from the 0.5 Mpc aperture; two faint sources were excised from the 1 Mpc aperture and an additional three faint sources from the 1.5 Mpc aperture.

A2244: The cluster looks regular but this may be a result of the low $S/N$ of the observation. No sources needed to be removed from the 0.5 Mpc aperture; one bright and one faint source were removed from the 1 Mpc aperture and another faint source from the 1.5 Mpc aperture.

A2255: This cluster does not have a large central emission peak and instead appears to have a large core. No sources needed to be removed from the 0.5 Mpc aperture. Four faint sources were removed from the 1 Mpc aperture. Four additional faint sources were removed from the 1.5 Mpc aperture

A2256: There are six observations of A2256 carefully placed at various positions to maximize coverage of the cluster within the ring of the PSPC; the ROSAT sequence numbers (exposure times) of these pointings are wp100110 (17865s), wp800162 (9108s), wp800163 (10803s), wp800339 (4978s), wp800340 (9430s), and wp800341 (10480s). wp100110 is roughly pointed on the cluster center, but it is displaced about $5'$ South of the field center. wp800162 and wp800339 are identical pointings where the center is pointed on the PSPC ring and the region NW of the cluster is centered on the field. The remaining pointings are symmetrically placed in a similar manner so that the cluster center is on the ring. The careful placement of the observations allows the 1.5 Mpc aperture to be enclosed by exposures entirely within the PSPC ring.

There are indeed many sources in this merged field as has been reported by Henry, Briel, & Nulsen (1993), most of which appear to lie at large distances from the cluster center. Within the 0.5 Mpc aperture ($\sim 8'$ radius) there are no obvious sources above the cluster continuum. 5 bright sources and seven faint sources were removed from the 1 Mpc aperture and three more bright and 11 faint sources were removed from the 1.5 Mpc aperture.

Note that the power ratios are generally quite large and indicative of an unrelaxed cluster for the 0.5 aperture, but at 1 and 1.5 Mpc the ratios suggest a near equilibrium state. Also, the 0.5 aperture power ratios are nearly the largest of our sample and suggest that A2256 is not so *typical* a cluster as has been suggested.

A2319: Two observations (exposure times: 3171s and 1505s) pointed on the same coordinates were merged into one observation. Apart from the cluster itself, there is only one faint source within the ring and hence we resorted to using the cluster center and the faint source to register the images to a common coordinate frame; the required shifts between the image were about one-half of a pixel and consistent with no shift. The aforementioned faint source was removed from the 1 Mpc aperture.

A2382: This cluster is quite irregular with several distinct emission peaks. Two observations (exposure times: 17444s and 8231s) pointed on the same coordinates were merged into one observation; several bright sources in the field were used for registration of the images to a reference coordinate frame. Two sources were removed from within the 0.5 Mpc and three faint sources straddling the aperture. Eight faint sources were removed from the 1 Mpc aperture. Since the cluster center is displaced to the South of the field center, the 1.5 Mpc aperture does not fit within the PSPC ring.

A2589: Two bright sources were removed from the 0.5 Mpc aperture of this mostly smooth-looking cluster.

A2597: This is a very symmetrical and smooth cluster. No obvious sources needed to be removed from the 0.5 Mpc aperture. Two bright and two faint sources straddling the 1 Mpc aperture were removed. One additional faint source was removed from the 1.5 Mpc aperture.

A2634: We did not include this cluster in the sample because the emission is contaminated by a background galaxy cluster that extends over several arcminutes.

A2657: Although for the most part smooth in appearance, there is some indication of bimodality in the inner arcminute. Two bright and seven faint sources were removed from the 0.5 Mpc aperture.

A2670: Two emission peaks separated by 1.5′ are resolved by the PSPC suggesting that A2670 is indeed undergoing a major merger. Although no sources needed to be removed from the 0.5 Mpc aperture, two bright and six faint sources were excised from the 1 Mpc aperture. Another three bright and two faint sources were removed from the 1.5 Mpc aperture. Part of an extended source on the edge of the ring to the East also was also removed. The nature of this extended source is unknown to us but if it is shown to be a subcluster associated with A2670 then its emission should be included in the aperture.

A2717: This faint cluster appears essentially regular. We removed 1 faint source in the 0.5 Mpc aperture and 8 faint sources in the 1 Mpc aperture.

A3158: The low $S/N$ of this observation does not allow any strong conclusions about its structure to be drawn. We used a low-exposure (3022s) observation during the PV phase that is not well centered on the cluster; only the 0.5 Mpc aperture fits within the ring. No sources were removed.

A3532: This is a PV-phase observation roughly centered on A3532 and includes A3530 (which is included in Ebeling's catalogue) which is located on the ring to the West. We do not attempt to remove A3530 and thus assume that it is gravitationally associated with A3532 and not a chance coincidence; the contribution to the X-ray emission only affects the power ratios in the 1 Mpc aperture. No sources were removed from the 0.5 Mpc aperture, but two were removed from the 1 Mpc aperture.

A3558: The dominant component of the cluster is mostly regular but with a small center displacement. There is substantial structure in this image to the East of the center of A3558. An obvious bridge of emission connects A3558 to two clumps just outside the ring. Using NED we identified one of these objects as the galaxy cluster AM 1328-313 which is at redshift 0.04380, essentially the same as A3558 – they are both members of the Shapley Supercluster. Two sources needed to be removed within the 0.5 Mpc aperture and an additional seven were removed from the 1 Mpc aperture.

A3562: The isophotes appear quite distorted and certainly not elliptical. The cluster AM 1328-313 lies to the SW just outside the ring in this image and A3558 is visible at the far W edge of the field. Two sources were removed from the 0.5 Mpc aperture and an additional 16 were removed for the 1 Mpc aperture.

A3667: This cluster has highly elongated and distorted isophotes. Four bright sources were removed within the ring including one near the center of the cluster.

A3921: This irregular cluster appears to be undergoing a merger of a subclump located about 1 Mpc to the West. No sources were removed from the 0.5 Mpc aperture. Two bright and four faint sources were removed from the 1 Mpc aperture. An additional four bright and five faint sources were removed from the 1.5 Mpc aperture.

A4038: Also known as Klemola44, this essentially elliptical cluster has two sources within the 0.5 Mpc aperture. There is a faint, extended source $10'$ to the N-E which coincides with between 5-10 galaxies in A4038 (Green, Godwin, & Peach 1990) so it may represent emission from a group within the cluster – hence we do not remove it. The other source does not obviously correspond to galaxies in A4038 and thus we do remove it.

A4059: This is a very smooth, single-component cluster. There is only one fairly bright point source within the central ring that was removed and was located in the 1 Mpc aperture only.

CYGNUS-A: This is a highly irregular cluster which has a roughly circular core region and a thick, bright emission tail that extends several arcminutes to the NW. Only one faint source was removed from the 0.5 Mpc aperture.

HYDRA-A: This cluster is centrally condensed and quite regular in appearance. We removed 1 bright and three faint sources within the 0.5 Mpc aperture and an additional four bright and six faint sources from the 1 Mpc aperture.

MKW3s: This poor cluster looks mostly smooth but quite elongated within the 0.5 Mpc aperture. No sources needed to be removed.

OPHIUCHUS: One bright source within the 0.5 Mpc aperture was removed. The isophotes of this cluster are somewhat asymmetrical. Also, the background rate is high because of the low Galactic latitude.

TRIANGULUM AUSTRALIS: The bright central emission peak appears to be slightly offset from the outer isophotes. Two faint sources were removed from the 1 Mpc aperture (none within 0.5 Mpc).

Table 1: The Sample

| NAME | 0.1-2.4 keV Flux ($10^{-12}$ erg cm$^{-2}$ s$^{-1}$) | Redshift | Exposure (s) | Background ($10^{-4}$ cts s$^{-1}$ arcmin$^{-2}$) | BM Type |
|---|---|---|---|---|---|
| OPH. | 345.00 | 0.0280 | 3932 | 7.36 | ... |
| A1656 | 340.37 | 0.0232 | ... | 4.08 | II |
| TRI-AUST | 104.00 | 0.0510 | 7338 | 4.30 | ... |
| A2319 | 101.00 | 0.0564 | 4676 | 7.56 | II-III |
| A2199 | 100.30 | 0.0299 | 10563 | 3.38 | I |
| A496 | 83.17 | 0.0327 | 8972 | 4.61 | I: |
| A3667 | 82.89 | 0.0530 | 12560 | 5.10 | I-II |
| A85 | 80.61 | 0.0556 | 10240 | 3.22 | I |
| A1795 | 72.42 | 0.0622 | 36829 | 3.03 | I |
| A754 | 67.84 | 0.0534 | 6359 | 3.21 | I-II: |
| A3558 | 66.85 | 0.0478 | 30213 | 4.89 | I |
| A2029 | 66.67 | 0.0768 | 12550 | 5.50 | I |
| A2142 | 65.22 | 0.0899 | 18873 | 2.65 | II |
| A4038 | 56.83 | 0.0283 | 3353 | 3.29 | III |
| A2256 | 54.38 | 0.0581 | ... | 2.50 | II-III: |
| A3266 | 53.44 | 0.0594 | 13560 | 3.09 | I-II |
| A2052 | 53.01 | 0.0348 | 9247 | 6.25 | I-II |
| HYDRA-A | 45.30 | 0.0522 | 18403 | 2.43 | I |
| A401 | 45.00 | 0.0748 | 14262 | 2.07 | I |
| A478 | 42.81 | 0.0881 | 22139 | 1.71 | ... |
| A2063 | 42.68 | 0.0355 | 10198 | 7.24 | II: |
| A119 | 42.12 | 0.0440 | 15203 | 3.33 | II-III |
| A644 | 39.20 | 0.0704 | 10285 | 2.29 | III: |
| A3158 | 39.19 | 0.0590 | 3022 | 2.95 | I-II |
| CYGNUS-A | 38.30 | 0.0561 | 6270 | 7.48 | I |
| A4059 | 34.88 | 0.0478 | 5514 | 2.79 | I |
| MKW3S | 34.70 | 0.0430 | 9996 | 6.24 | II-III |
| A3562 | 34.47 | 0.0499 | 20202 | 3.99 | I |
| A1651 | 29.05 | 0.0845 | 7435 | 3.55 | I-II |
| A2589 | 28.61 | 0.0415 | 7293 | 2.37 | I |
| A2597 | 27.38 | 0.0852 | 7243 | 2.25 | III |
| A2657 | 25.23 | 0.0414 | 18911 | 2.25 | III |
| A2204 | 24.77 | 0.1523 | 5359 | 8.47 | II |
| A2244 | 24.71 | 0.0968 | 2965 | 2.39 | I-II: |
| A3532 | 24.60 | 0.0585 | 8620 | 3.36 | II-III |
| A2163 | 20.70 | 0.2010 | 10525 | 3.98 | ... |
| A400 | 19.31 | 0.0238 | 23615 | 2.23 | II-III |
| A2255 | 19.02 | 0.0808 | 14555 | 2.29 | II-III: |
| A2107 | 18.05 | 0.0421 | 8274 | 3.81 | I |
| A3921 | 16.82 | 0.0960 | 12007 | 3.09 | II |
| A1914 | 16.37 | 0.1712 | 9040 | 2.72 | II: |
| A1689 | 16.26 | 0.1832 | 13957 | 3.14 | II-III: |
| A1413 | 16.22 | 0.1427 | 7798 | 2.52 | I |
| A1991 | 14.69 | 0.0579 | 21281 | 4.48 | I: |
| A1750 | 14.62 | 0.0855 | 13148 | 2.79 | II-III: |
| A2034 | 13.93 | 0.1510 | 8958 | 2.90 | II-III: |
| A665 | 13.71 | 0.1819 | 38641 | 3.17 | III: |
| A2670 | 12.31 | 0.0761 | 17701 | 2.95 | I-II |
| A2717 | 10.90 | 0.0498 | 9907 | 2.31 | I-II |
| A21 | 10.86 | 0.0946 | 8680 | 3.67 | I: |
| A545 | 10.63 | 0.1540 | 14285 | 2.05 | III |
| A1068 | 10.52 | 0.1386 | 10648 | 2.59 | I |
| A586 | 9.93 | 0.1710 | 4082 | 2.04 | I |
| A1837 | 9.17 | 0.0376 | 15727 | 3.38 | I-II |
| A2218 | 8.74 | 0.1710 | 44530 | 3.49 | II: |
| A500 | 8.45 | 0.0666 | 18400 | 2.13 | III |
| A1361 | 7.12 | 0.1167 | 5675 | 2.08 | I-II |
| A2382 | 6.37 | 0.0648 | 25675 | 3.32 | II-III |
| A514 | 5.00 | 0.0731 | 18111 | 2.24 | II-III: |

Note. — Fluxes are from Ebeling (1993) otherwise from this paper (see §3). Redshifts compiled from Struble & Rood (1991), Edge et al. (1990), and NED. BM Types are primarily from Leir & van den Bergh (1977), otherwise Bautz & Morgan (1970), Sandage & Hardy (1973), Bahcall (1980), and Abell, Corwin, & Olowen (1989).

Table 2.  Power Ratios

| NAME | $R_{ap}$ | $P_1^{(pk)}/P_0^{(pk)}$ | | $P_2/P_0$ | | $P_3/P_0$ | | $P_4/P_0$ | |
|---|---|---|---|---|---|---|---|---|---|
| OPH. | 0.50 | 53. | 39. - 144. | 9.4 | 6.8 - 11.3 | 1.971 | 1.371 - 2.571 | 0.023 | 0.005 - 0.084 |
| A1656 | 0.45 | 237. | 173. - 317. | 42.5 | 41.3 - 45.6 | 0.270 | 0.141 - 0.342 | 0.172 | 0.112 - 0.221 |
| TRI-AUST | 0.50 | 483. | 223. - 559. | 41.5 | 33.7 - 49.4 | 0.234 | 0.083 - 0.832 | 0.222 | 0.068 - 0.499 |
| TRI-AUST | 1.00 | 93. | 40. - 122. | 9.8 | 6.9 - 11.3 | 0.217 | 0.052 - 0.460 | 0.057 | 0.015 - 0.167 |
| A2319 | 0.50 | 1404. | 1213. - 1811. | 28.5 | 19.4 - 37.0 | 1.423 | 0.467 - 2.528 | 0.951 | 0.475 - 1.358 |
| A2319 | 1.00 | 721. | 630. - 917. | 3.2 | 1.3 - 4.6 | 0.820 | 0.436 - 1.542 | 0.023 | 0.003 - 0.106 |
| A2199 | 0.50 | 14. | 7. - 21. | 5.9 | 4.4 - 6.9 | 0.051 | 0.016 - 0.175 | 0.043 | 0.011 - 0.091 |
| A496 | 0.50 | 55. | 41. - 74. | 4.9 | 2.3 - 6.0 | 0.081 | 0.020 - 0.301 | 0.023 | 0.001 - 0.117 |
| A3667 | 0.50 | 322. | 98. - 656. | 84.2 | 74.0 - 95.2 | 1.965 | 1.110 - 3.072 | 0.289 | 0.085 - 0.555 |
| A3667 | 1.00 | 1042. | 784. - 1399. | 84.8 | 75.4 - 94.9 | 1.818 | 1.100 - 2.470 | 1.280 | 0.817 - 1.490 |
| A85 | 0.50 | 311. | 110. - 344. | 15.4 | 12.3 - 19.9 | 1.100 | 0.680 - 1.535 | 0.032 | 0.011 - 0.123 |
| A85 | 1.00 | 333. | 187. - 375. | 13.8 | 10.1 - 14.9 | 0.811 | 0.540 - 1.246 | 0.166 | 0.110 - 0.304 |
| A1795 | 0.50 | 259. | 27. - 267. | 12.8 | 11.8 - 14.3 | 0.003 | 0.000 - 0.035 | 0.004 | 0.000 - 0.020 |
| A1795 | 1.00 | 77. | 18. - 88. | 2.0 | 1.4 - 2.3 | 0.037 | 0.014 - 0.086 | 0.001 | 0.000 - 0.014 |
| A754 | 0.50 | 5704. | 5065. - 8054. | 270.8 | 246.3 - 305.3 | 2.062 | 1.213 - 3.384 | 3.196 | 1.844 - 3.951 |
| A3558 | 0.50 | 936. | 875. - 998. | 46.7 | 40.2 - 50.2 | 0.184 | 0.093 - 0.350 | 0.164 | 0.063 - 0.253 |
| A3558 | 0.97 | 325. | 299. - 360. | 52.0 | 49.2 - 58.6 | 3.210 | 2.758 - 3.876 | 0.451 | 0.282 - 0.570 |
| A2029 | 0.50 | 37. | 28. - 53. | 14.0 | 11.1 - 17.0 | 0.031 | 0.002 - 0.098 | 0.050 | 0.018 - 0.153 |
| A2029 | 1.00 | 9. | 5. - 13. | 1.7 | 1.2 - 2.5 | 0.004 | 0.002 - 0.061 | 0.073 | 0.032 - 0.146 |
| A2029 | 1.50 | 5. | 2. - 8. | 2.0 | 0.7 - 2.1 | 0.020 | 0.002 - 0.094 | 0.060 | 0.033 - 0.145 |
| A2142 | 0.50 | 10. | 6. - 448. | 87.9 | 76.6 - 92.8 | 0.393 | 0.118 - 0.700 | 1.073 | 0.627 - 1.217 |
| A2142 | 1.00 | 70. | 61. - 348. | 30.0 | 26.3 - 32.9 | 0.549 | 0.381 - 0.752 | 0.344 | 0.275 - 0.447 |
| A2142 | 1.50 | 27. | 20. - 132. | 5.2 | 4.4 - 6.4 | 0.090 | 0.028 - 0.178 | 0.047 | 0.014 - 0.095 |
| A4038 | 0.50 | 14. | 1. - 40. | 32.9 | 26.8 - 43.7 | 0.607 | 0.163 - 1.453 | 0.106 | 0.019 - 0.365 |
| A2256 | 0.50 | 3844. | 2921. - 3937. | 102.6 | 94.9 - 107.0 | 1.582 | 1.198 - 2.168 | 0.395 | 0.213 - 0.551 |
| A2256 | 1.00 | 2572. | 2101. - 2619. | 20.8 | 18.8 - 23.7 | 0.065 | 0.026 - 0.178 | 0.151 | 0.081 - 0.219 |
| A2256 | 1.50 | 1213. | 999. - 1276. | 4.8 | 4.1 - 7.1 | 0.144 | 0.062 - 0.431 | 0.143 | 0.062 - 0.230 |
| A3266 | 0.50 | 3726. | 2965. - 3996. | 39.0 | 32.0 - 47.0 | 0.266 | 0.059 - 1.071 | 0.319 | 0.103 - 0.543 |
| A3266 | 0.95 | 1907. | 1513. - 2071. | 24.7 | 23.1 - 31.6 | 0.177 | 0.037 - 0.468 | 0.086 | 0.024 - 0.179 |
| A2052 | 0.50 | 7. | 2. - 16. | 13.0 | 8.4 - 17.2 | 0.033 | 0.006 - 0.296 | 0.031 | 0.002 - 0.163 |
| HYDRA-A | 0.50 | 21. | 15. - 29. | 4.9 | 3.8 - 6.2 | 0.044 | 0.008 - 0.144 | 0.029 | 0.008 - 0.082 |
| HYDRA-A | 1.00 | 38. | 32. - 48. | 5.1 | 3.8 - 6.3 | 0.133 | 0.044 - 0.279 | 0.055 | 0.018 - 0.163 |
| A401 | 0.50 | 115. | 4. - 154. | 38.6 | 31.4 - 46.7 | 0.460 | 0.213 - 0.997 | 0.157 | 0.042 - 0.359 |
| A401 | 1.00 | 193. | 120. - 261. | 8.7 | 6.2 - 10.9 | 0.098 | 0.012 - 0.249 | 0.045 | 0.015 - 0.147 |
| A401 | 1.50 | 239. | 173. - 297. | 7.1 | 4.4 - 8.5 | 0.273 | 0.073 - 0.521 | 0.002 | 0.001 - 0.066 |
| A478 | 0.50 | 25. | 14. - 34. | 13.5 | 10.8 - 16.5 | 0.099 | 0.014 - 0.244 | 0.025 | 0.012 - 0.113 |
| A478 | 1.00 | 5. | 2. - 8. | 2.2 | 1.5 - 3.1 | 0.045 | 0.007 - 0.133 | 0.001 | 0.000 - 0.020 |
| A478 | 1.50 | 1. | 0. - 3. | 0.6 | 0.2 - 1.2 | 0.003 | 0.001 - 0.068 | 0.002 | 0.001 - 0.045 |
| A2063 | 0.50 | 66. | 33. - 193. | 10.0 | 6.7 - 15.3 | 0.196 | 0.044 - 0.746 | 0.076 | 0.004 - 0.268 |
| A119 | 0.50 | 128. | 38. - 366. | 21.4 | 14.6 - 29.9 | 9.847 | 6.942 - 12.276 | 1.218 | 0.651 - 1.918 |
| A119 | 0.94 | 234. | 137. - 425. | 5.3 | 3.0 - 8.2 | 2.084 | 1.228 - 3.694 | 0.155 | 0.073 - 0.542 |
| A644 | 0.50 | 893. | 787. - 981. | 36.1 | 30.5 - 43.1 | 0.521 | 0.171 - 0.926 | 0.030 | 0.004 - 0.129 |
| A644 | 1.00 | 419. | 376. - 480. | 7.0 | 5.1 - 10.0 | 0.097 | 0.010 - 0.315 | 0.030 | 0.003 - 0.125 |
| A644 | 1.45 | 200. | 176. - 250. | 2.2 | 0.6 - 3.1 | 0.042 | 0.006 - 0.366 | 0.205 | 0.157 - 0.529 |
| A3158 | 0.50 | 15. | 2. - 226. | 43.8 | 26.3 - 63.2 | 0.396 | 0.050 - 2.186 | 0.294 | 0.027 - 1.087 |
| CYG-A | 0.50 | 977. | 780. - 1093. | 30.8 | 19.9 - 39.4 | 1.030 | 0.377 - 2.201 | 0.058 | 0.008 - 0.245 |
| CYG-A | 1.00 | 2321. | 2151. - 2623. | 63.3 | 51.6 - 77.6 | 3.522 | 2.126 - 5.344 | 2.065 | 1.137 - 3.126 |
| A4059 | 0.50 | 38. | 16. - 78. | 23.6 | 17.1 - 33.5 | 0.187 | 0.038 - 0.687 | 0.239 | 0.026 - 0.592 |
| A4059 | 1.00 | 26. | 8. - 48. | 6.7 | 2.2 - 9.4 | 0.269 | 0.023 - 0.930 | 0.092 | 0.045 - 0.715 |
| MKW3s | 0.50 | 100. | 71. - 150. | 16.0 | 10.4 - 20.9 | 0.053 | 0.009 - 0.380 | 0.074 | 0.011 - 0.276 |
| A3562 | 0.50 | 21. | 7. - 76. | 4.7 | 1.9 - 8.4 | 5.645 | 4.051 - 7.449 | 0.367 | 0.137 - 0.702 |
| A3562 | 1.00 | 123. | 85. - 184. | 32.6 | 33.8 - 49.4 | 1.052 | 0.580 - 1.878 | 0.247 | 0.102 - 0.707 |
| A2589 | 0.50 | 47. | 19. - 94. | 26.5 | 16.3 - 34.3 | 0.107 | 0.020 - 0.694 | 0.026 | 0.009 - 0.367 |
| A1651 | 0.50 | 87. | 45. - 140. | 9.9 | 5.5 - 17.5 | 0.019 | 0.006 - 0.688 | 0.205 | 0.028 - 0.501 |
| A1651 | 1.00 | 52. | 30. - 87. | 10.5 | 6.4 - 15.2 | 0.145 | 0.031 - 0.720 | 0.059 | 0.005 - 0.249 |
| A1651 | 1.50 | 22. | 9. - 47. | 4.4 | 3.2 - 9.8 | 0.272 | 0.037 - 0.922 | 0.111 | 0.021 - 0.307 |
| A2597 | 0.50 | 6. | 1. - 16. | 10.5 | 6.6 - 14.0 | 0.016 | 0.003 - 0.325 | 0.074 | 0.005 - 0.169 |
| A2597 | 1.00 | 1. | 0. - 9. | 3.2 | 1.5 - 4.9 | 0.437 | 0.121 - 1.004 | 0.010 | 0.004 - 0.116 |
| A2597 | 1.50 | 7. | 2. - 21. | 1.6 | 0.3 - 3.7 | 0.332 | 0.061 - 0.963 | 0.129 | 0.013 - 0.335 |
| A2657 | 0.50 | 28. | 13. - 48. | 29.7 | 26.3 - 39.2 | 0.495 | 0.164 - 1.407 | 0.295 | 0.093 - 0.728 |

Table 2—Continued

| NAME | $R_{ap}$ | $P_1^{(pk)}/P_0^{(pk)}$ | | $P_2/P_0$ | | $P_3/P_0$ | | $P_4/P_0$ | |
|---|---|---|---|---|---|---|---|---|---|
| A2204 | 0.50 | 14. | 5. - 42. | 2.7 | 1.0 - 5.9 | 0.075 | 0.011 - 0.655 | 0.157 | 0.030 - 0.427 |
| A2204 | 1.00 | 9. | 1. - 19. | 1.0 | 0.3 - 2.9 | 0.453 | 0.146 - 0.884 | 0.040 | 0.006 - 0.206 |
| A2204 | 1.50 | 1. | 0. - 10. | 0.0 | 0.0 - 0.9 | 0.050 | 0.004 - 0.512 | 0.148 | 0.030 - 0.527 |
| A2244 | 0.50 | 94. | 51. - 196. | 3.1 | 0.3 - 10.3 | 0.261 | 0.018 - 1.860 | 0.109 | 0.020 - 0.688 |
| A2244 | 1.00 | 14. | 3. - 61. | 7.1 | 3.3 - 14.3 | 0.532 | 0.050 - 1.786 | 0.026 | 0.006 - 0.461 |
| A2244 | 1.50 | 13. | 1. - 60. | 0.7 | 0.1 - 3.3 | 0.877 | 0.276 - 3.041 | 0.616 | 0.252 - 1.734 |
| A3532 | 0.50 | 36. | 16. - 233. | 20.3 | 10.4 - 31.1 | 0.344 | 0.067 - 2.317 | 0.110 | 0.017 - 0.927 |
| A3532 | 1.00 | 91. | 50. - 286. | 25.1 | 18.6 - 50.0 | 2.323 | 0.745 - 5.280 | 0.728 | 0.232 - 2.177 |
| A2163 | 0.50 | 2172. | 173. - 2427. | 17.9 | 4.5 - 30.0 | 1.797 | 0.212 - 4.208 | 0.913 | 0.064 - 1.791 |
| A2163 | 1.00 | 1993. | 494. - 2154. | 7.9 | 4.3 - 15.1 | 0.156 | 0.014 - 0.728 | 0.709 | 0.238 - 1.135 |
| A2163 | 1.50 | 1284. | 400. - 1400. | 4.6 | 1.7 - 9.3 | 0.902 | 0.320 - 2.080 | 0.662 | 0.349 - 1.379 |
| A400 | 0.46 | 465. | 258. - 607. | 49.7 | 26.8 - 52.5 | 4.616 | 2.917 - 7.669 | 0.366 | 0.194 - 1.356 |
| A2255 | 0.50 | 786. | 144. - 2333. | 25.8 | 12.7 - 35.5 | 0.054 | 0.022 - 1.171 | 0.223 | 0.010 - 0.942 |
| A2255 | 1.00 | 587. | 131. - 1549. | 18.4 | 13.2 - 26.9 | 0.455 | 0.075 - 1.179 | 0.034 | 0.009 - 0.282 |
| A2255 | 1.50 | 135. | 21. - 448. | 2.5 | 2.5 - 9.7 | 0.394 | 0.090 - 1.055 | 0.619 | 0.311 - 1.387 |
| A2107 | 0.50 | 15. | 1. - 44. | 6.8 | 4.0 - 17.2 | 0.583 | 0.106 - 2.068 | 0.102 | 0.015 - 0.597 |
| A3921 | 0.50 | 97. | 18. - 305. | 63.4 | 46.7 - 83.7 | 0.705 | 0.039 - 2.206 | 0.360 | 0.066 - 1.012 |
| A3921 | 1.00 | 508. | 228. - 632. | 135.7 | 115.2 - 155.3 | 12.720 | 9.281 - 15.788 | 2.309 | 1.325 - 3.624 |
| A3921 | 1.50 | 461. | 252. - 572. | 56.5 | 51.5 - 80.2 | 0.524 | 0.151 - 1.415 | 0.082 | 0.016 - 0.497 |
| A1914 | 0.50 | 631. | 8. - 808. | 17.6 | 9.5 - 24.2 | 1.981 | 0.740 - 3.500 | 0.072 | 0.013 - 0.389 |
| A1914 | 1.00 | 89. | 28. - 138. | 1.3 | 0.4 - 2.9 | 0.469 | 0.152 - 1.128 | 0.139 | 0.027 - 0.347 |
| A1914 | 1.50 | 35. | 12. - 61. | 2.0 | 0.5 - 3.6 | 0.232 | 0.033 - 0.710 | 0.022 | 0.004 - 0.143 |
| A1689 | 0.50 | 521. | 253. - 552. | 10.4 | 4.7 - 14.4 | 0.213 | 0.018 - 0.699 | 0.054 | 0.005 - 0.302 |
| A1689 | 1.00 | 193. | 34. - 226. | 6.5 | 4.2 - 8.9 | 0.016 | 0.006 - 0.230 | 0.033 | 0.003 - 0.112 |
| A1689 | 1.50 | 108. | 12. - 128. | 2.5 | 0.9 - 3.5 | 0.037 | 0.006 - 0.223 | 0.018 | 0.001 - 0.101 |
| A1413 | 0.50 | 19. | 3. - 429. | 52.2 | 35.1 - 75.0 | 0.022 | 0.009 - 1.085 | 0.112 | 0.049 - 1.143 |
| A1413 | 1.00 | 40. | 11. - 73. | 25.4 | 18.3 - 33.2 | 0.010 | 0.009 - 0.409 | 0.120 | 0.020 - 0.382 |
| A1413 | 1.50 | 27. | 2. - 54. | 7.3 | 2.5 - 9.1 | 0.026 | 0.008 - 0.547 | 0.019 | 0.012 - 0.308 |
| A1991 | 0.50 | 2. | 0. - 13. | 9.9 | 5.7 - 13.6 | 0.140 | 0.019 - 0.702 | 0.001 | 0.002 - 0.131 |
| A1991 | 1.00 | 16. | 4. - 44. | 1.2 | 0.2 - 3.9 | 0.642 | 0.095 - 1.955 | 0.036 | 0.017 - 0.588 |
| A1750 | 0.50 | 7. | 6. - 228. | 8.9 | 3.2 - 21.9 | 8.599 | 3.214 - 12.653 | 0.058 | 0.006 - 0.724 |
| A1750 | 1.00 | 3670. | 3076. - 4219. | 818.5 | 752.1 - 906.4 | 6.075 | 3.417 - 12.049 | 12.633 | 9.735 - 16.556 |
| A1750 | 1.40 | 3996. | 3805. - 5028. | 311.0 | 266.9 - 346.8 | 0.679 | 0.065 - 3.112 | 7.333 | 5.672 - 10.618 |
| A2034 | 0.50 | 1055. | 67. - 2944. | 30.6 | 13.3 - 51.2 | 0.442 | 0.136 - 3.538 | 0.449 | 0.018 - 1.303 |
| A2034 | 1.00 | 373. | 43. - 816. | 25.8 | 13.8 - 36.2 | 0.944 | 0.308 - 2.004 | 0.556 | 0.209 - 1.215 |
| A2034 | 1.50 | 377. | 64. - 676. | 16.4 | 8.7 - 20.8 | 0.306 | 0.071 - 1.160 | 0.019 | 0.006 - 0.337 |
| A665 | 0.50 | 3134. | 1393. - 3255. | 13.8 | 6.4 - 21.7 | 2.992 | 1.061 - 4.462 | 0.264 | 0.104 - 0.973 |
| A665 | 1.00 | 2399. | 1563. - 2542. | 1.2 | 0.2 - 2.7 | 1.254 | 0.843 - 2.324 | 0.475 | 0.185 - 0.772 |
| A665 | 1.50 | 1467. | 890. - 1598. | 0.3 | 0.1 - 1.8 | 0.501 | 0.208 - 1.037 | 0.017 | 0.003 - 0.131 |
| A2670 | 0.50 | 461. | 39. - 1033. | 0.6 | 0.1 - 4.7 | 1.758 | 0.880 - 3.502 | 0.021 | 0.003 - 0.298 |
| A2670 | 1.00 | 119. | 21. - 274. | 1.2 | 0.2 - 5.1 | 1.105 | 0.330 - 2.391 | 0.116 | 0.007 - 0.616 |
| A2670 | 1.50 | 81. | 23. - 203. | 1.8 | 1.6 - 17.6 | 0.800 | 0.136 - 2.959 | 0.104 | 0.035 - 0.942 |
| A2717 | 0.50 | 174. | 98. - 303. | 2.2 | 0.3 - 8.7 | 0.432 | 0.080 - 2.246 | 0.041 | 0.009 - 0.575 |
| A2717 | 0.95 | 15. | 1. - 48. | 3.0 | 0.5 - 14.0 | 1.340 | 0.308 - 6.574 | 0.052 | 0.036 - 1.784 |
| A21 | 0.50 | 294. | 16. - 927. | 73.4 | 52.1 - 119.2 | 4.443 | 0.750 - 9.033 | 0.293 | 0.029 - 1.518 |
| A21 | 1.00 | 6. | 1. - 232. | 53.1 | 29.3 - 75.6 | 0.893 | 0.050 - 2.774 | 0.267 | 0.030 - 1.078 |
| A21 | 1.50 | 33. | 21. - 207. | 11.6 | 1.3 - 17.0 | 1.030 | 0.123 - 4.204 | 0.622 | 0.083 - 2.395 |
| A545 | 0.50 | 452. | 233. - 1547. | 94.0 | 66.2 - 121.4 | 0.473 | 0.078 - 2.480 | 0.586 | 0.074 - 1.260 |
| A545 | 1.00 | 405. | 258. - 847. | 7.7 | 4.3 - 14.3 | 0.454 | 0.143 - 1.689 | 0.087 | 0.002 - 0.396 |
| A545 | 1.50 | 269. | 182. - 496. | 0.3 | 0.0 - 2.1 | 0.025 | 0.013 - 0.636 | 0.206 | 0.045 - 0.575 |
| A1068 | 0.50 | 383. | 144. - 486. | 17.4 | 10.5 - 21.9 | 0.137 | 0.011 - 0.673 | 0.144 | 0.007 - 0.319 |
| A1068 | 1.00 | 104. | 27. - 141. | 8.5 | 4.6 - 12.3 | 0.301 | 0.054 - 0.823 | 0.023 | 0.003 - 0.158 |
| A1068 | 1.50 | 49. | 7. - 84. | 2.9 | 0.7 - 5.6 | 0.027 | 0.008 - 0.611 | 0.046 | 0.003 - 0.265 |
| A586 | 0.50 | 53. | 10. - 947. | 0.4 | 0.4 - 9.8 | 1.505 | 0.228 - 5.527 | 0.275 | 0.025 - 1.346 |
| A586 | 1.00 | 62. | 14. - 207. | 5.3 | 0.6 - 16.8 | 0.147 | 0.044 - 1.374 | 0.281 | 0.016 - 1.119 |
| A586 | 1.50 | 37. | 9. - 139. | 1.2 | 0.3 - 11.2 | 0.096 | 0.037 - 2.055 | 0.163 | 0.013 - 1.426 |
| A1837 | 0.50 | 284. | 163. - 485. | 10.7 | 3.0 - 30.8 | 0.413 | 0.066 - 2.946 | 0.258 | 0.066 - 1.761 |
| A2218 | 0.50 | 738. | 68. - 863. | 30.6 | 18.8 - 34.7 | 0.103 | 0.008 - 0.540 | 0.009 | 0.002 - 0.201 |
| A2218 | 1.00 | 444. | 69. - 510. | 25.1 | 18.8 - 30.0 | 0.192 | 0.014 - 0.535 | 0.045 | 0.009 - 0.238 |
| A2218 | 1.50 | 280. | 37. - 323. | 13.3 | 11.3 - 22.9 | 0.159 | 0.036 - 0.653 | 0.002 | 0.005 - 0.211 |

Table 2—Continued

| NAME | $R_{ap}$ | $P_1^{(pk)}/P_0^{(pk)}$ | | $P_2/P_0$ | | $P_3/P_0$ | | $P_4/P_0$ | |
|------|----------|------|------|------|------|------|------|------|------|
| A500  | 0.50 | 389.  | 98. - 538.      | 17.9  | 8.6 - 40.5    | 0.095  | 0.048 - 1.720   | 1.197  | 0.181 - 2.297   |
| A500  | 1.00 | 637.  | 401. - 914.     | 10.3  | 4.2 - 25.8    | 0.515  | 0.063 - 2.391   | 0.396  | 0.019 - 1.344   |
| A1361 | 0.50 | 49.   | 15. - 159.      | 8.2   | 2.4 - 21.1    | 0.620  | 0.073 - 2.430   | 0.177  | 0.016 - 0.852   |
| A1361 | 1.00 | 12.   | 1. - 46.        | 8.7   | 3.0 - 19.0    | 0.641  | 0.027 - 2.271   | 0.050  | 0.014 - 0.861   |
| A1361 | 1.50 | 2.    | 1. - 61.        | 2.1   | 0.1 - 7.6     | 0.609  | 0.107 - 3.864   | 0.186  | 0.035 - 2.063   |
| A2382 | 0.50 | 1042. | 723. - 1662.    | 33.6  | 19.8 - 51.7   | 1.769  | 0.315 - 4.407   | 0.299  | 0.039 - 1.035   |
| A2382 | 1.00 | 496.  | 355. - 935.     | 47.2  | 19.5 - 54.7   | 0.668  | 0.175 - 3.617   | 0.945  | 0.380 - 2.530   |
| A514  | 0.50 | 9369. | 8344. - 10978.  | 273.2 | 225.0 - 329.2 | 22.712 | 13.847 - 32.811 | 0.679  | 0.043 - 2.292   |
| A514  | 1.00 | 4833. | 3728. - 5297.   | 299.6 | 242.0 - 358.4 | 5.825  | 2.574 - 11.480  | 13.619 | 10.415 - 17.973 |
| A514  | 1.45 | 2557. | 2148. - 4290.   | 179.2 | 177.4 - 271.6 | 7.396  | 2.562 - 13.764  | 0.155  | 0.044 - 1.595   |

Note. — The power ratios and their 90% confidence estimates are expressed in units of $10^{-7}$. $R_{ap}$ is the aperture radius in $h_{80}^{-1}$Mpc.

---

Fig. 1.—
ROSAT PSPC image for A1795 before (left) and after (right) excising point sources. For viewing purposes only we smoothed these images with a Gaussian ($\sigma = 15''$).

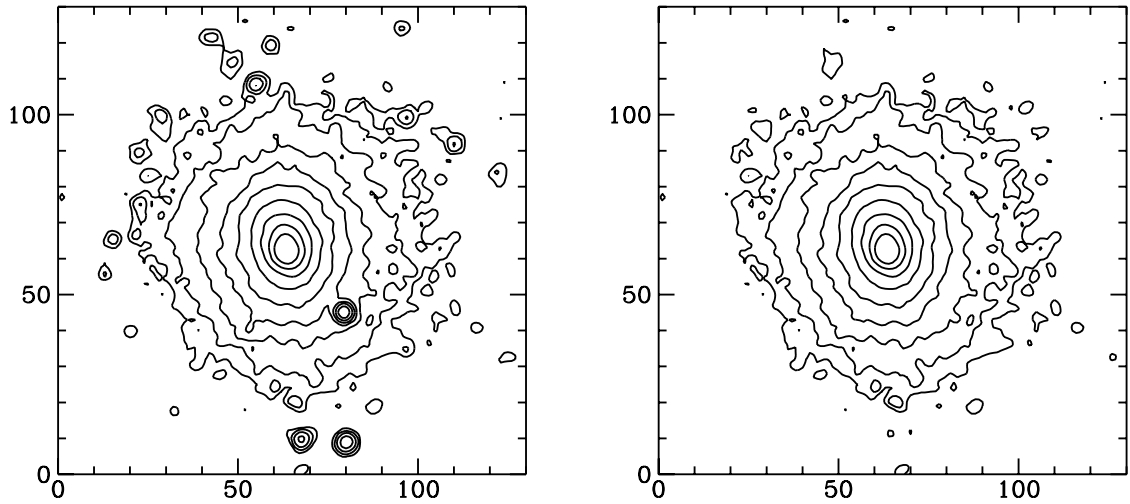

Fig. 2.—
Power ratios as a function of the number of excised sources for A1795. $P_1^{(pk)}/P_0^{(pk)}$ is denoted by crosses, $P_2/P_0$ by circles, $P_3/P_0$ by boxes, and $P_4/P_0$ by stars.

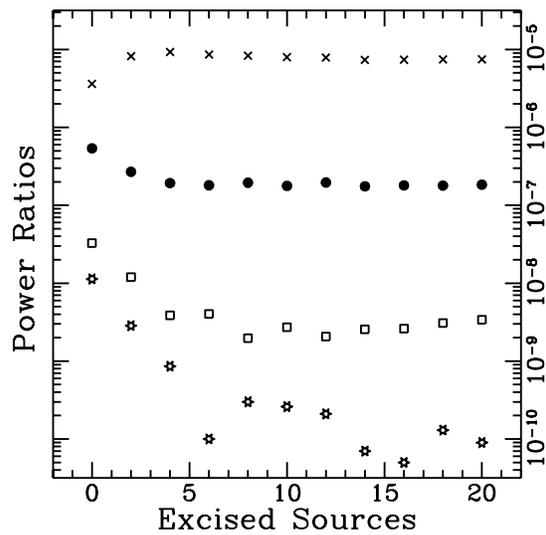

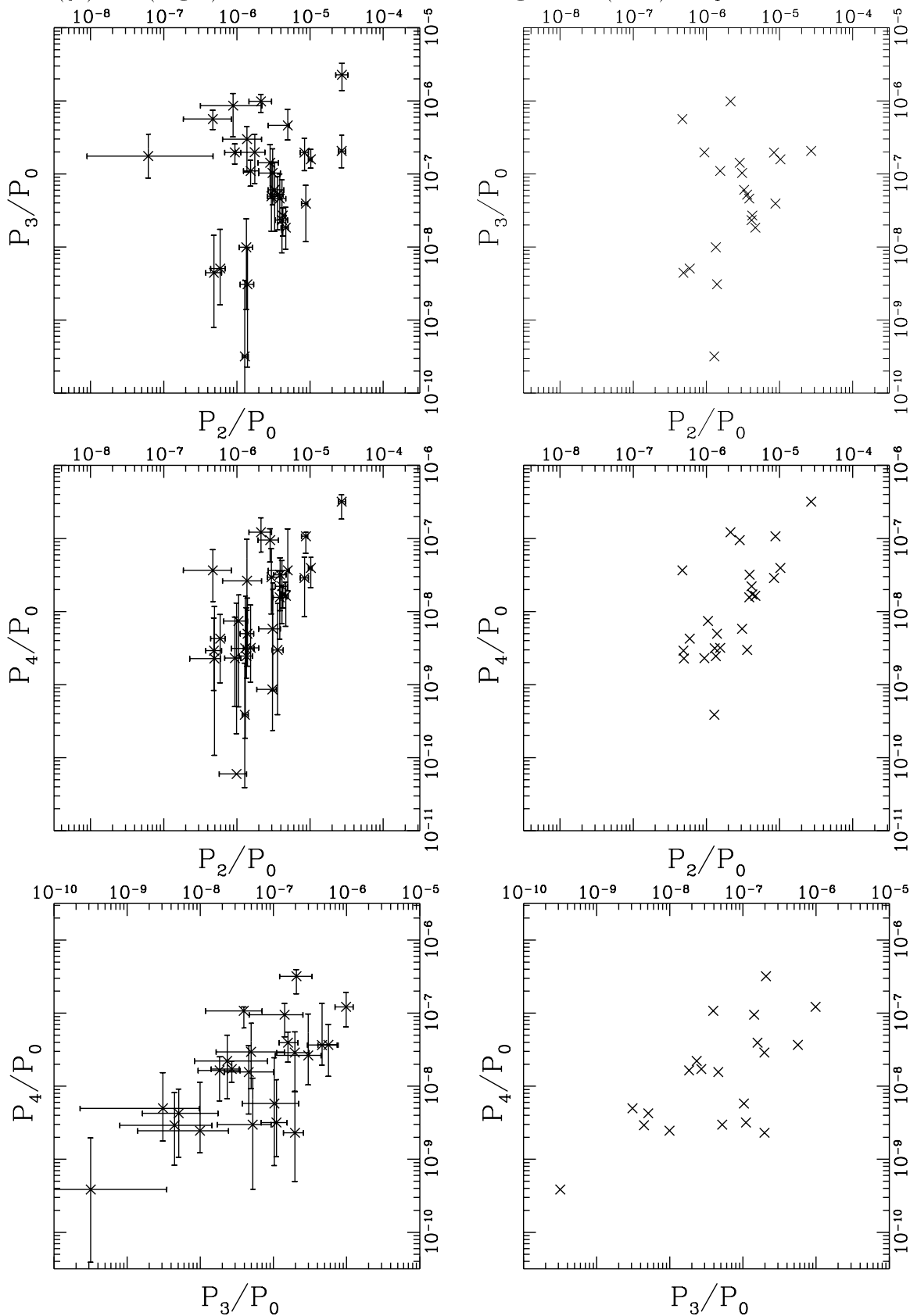

Fig. 3.—
(Left) Power-ratio correlations computed in the $0.5 h_{80}^{-1}$ Mpc aperture for the "best measured" clusters (§4) and (Right) those also included in the Edge et al. (1990) sample.

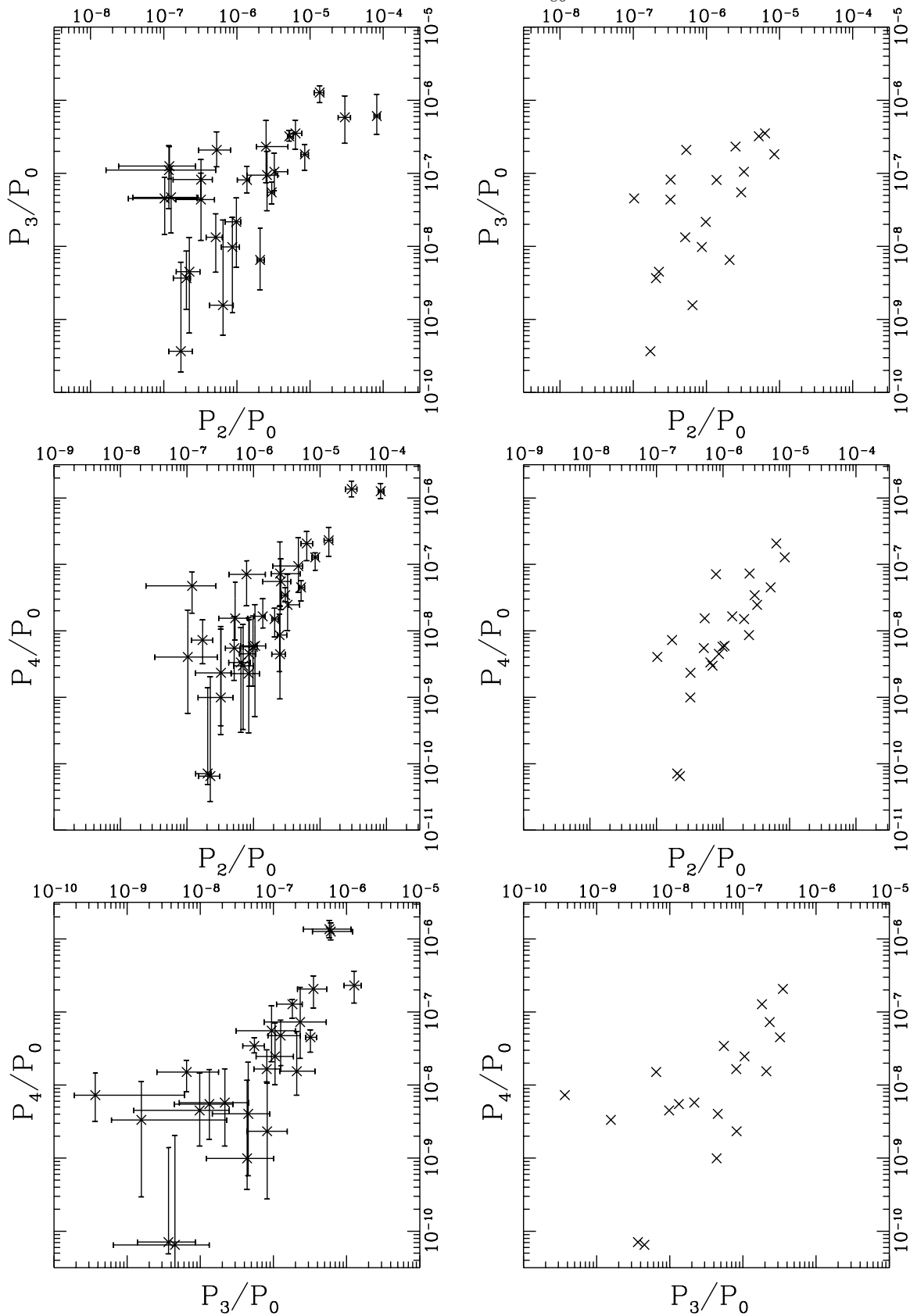

Fig. 4.—
Same as Figure 3 for the power ratios computed in the $1h_{80}^{-1}$ Mpc aperture.

Fig. 5.—
Power ratios for the "reference" clusters computed in the $0.5h_{80}^{-1}$ Mpc aperture (see §4).

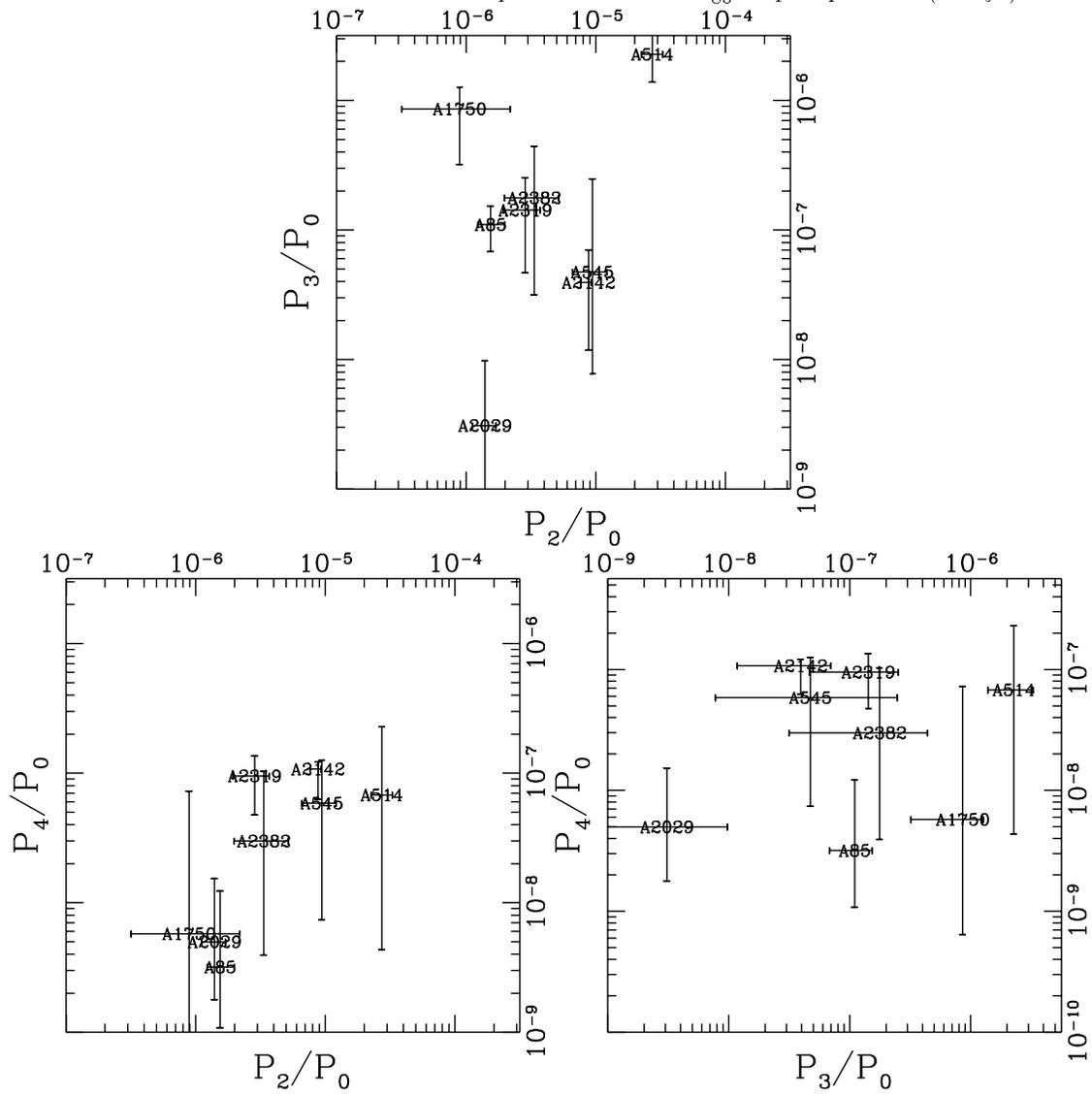

Fig. 6.—
Power ratios for the "reference" clusters computed in the $1h_{80}^{-1}$ Mpc aperture (see §4).

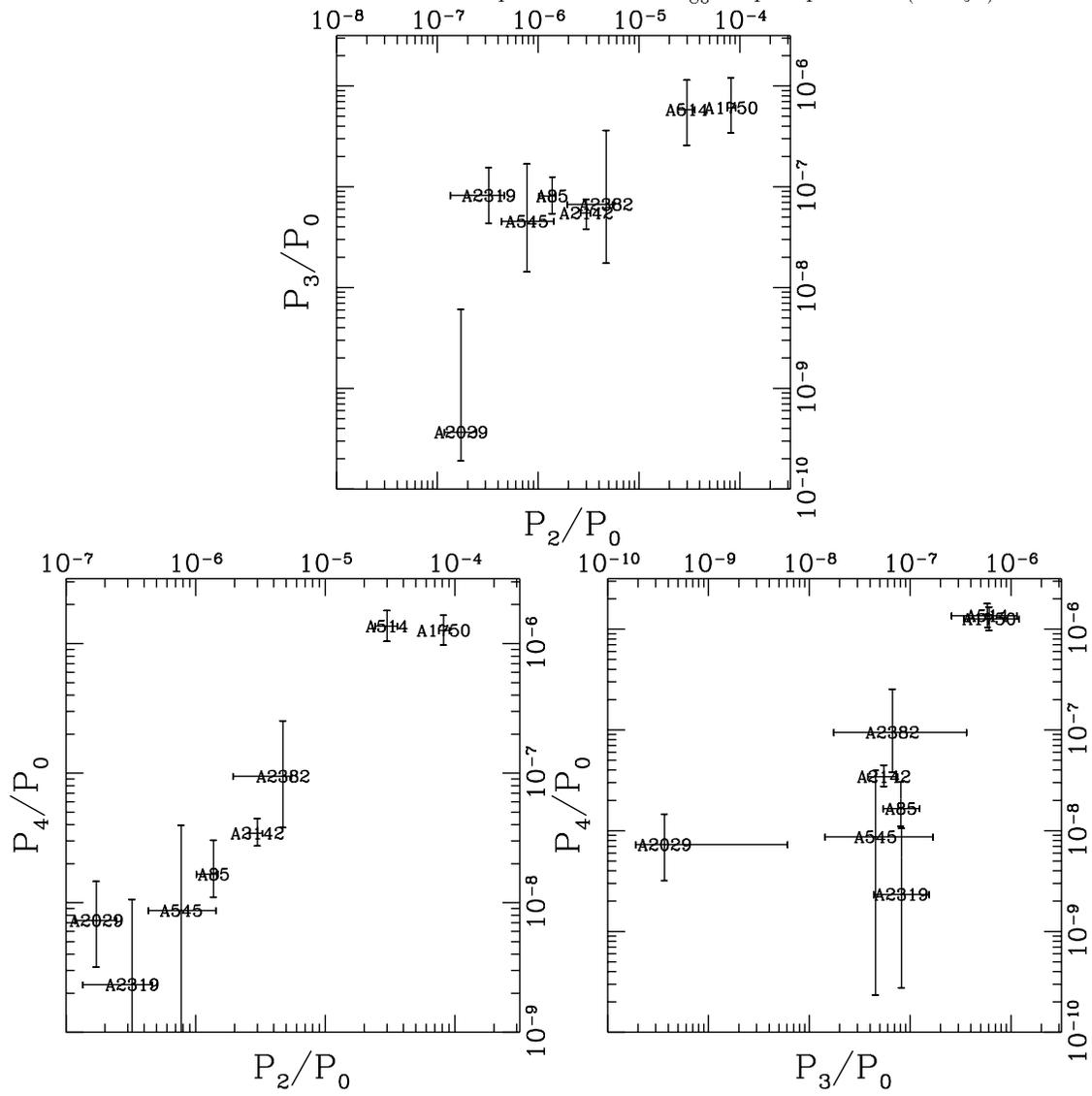

Fig. 7.—
Mass-flow rate from Fabian (1994) vs $P_2/P_0$ computed in the $1h_{80}^{-1}$ Mpc aperture for the clusters in our sample corresponding to that of Edge et al. (1990). The clusters on the bottom axis have no detected cooling.

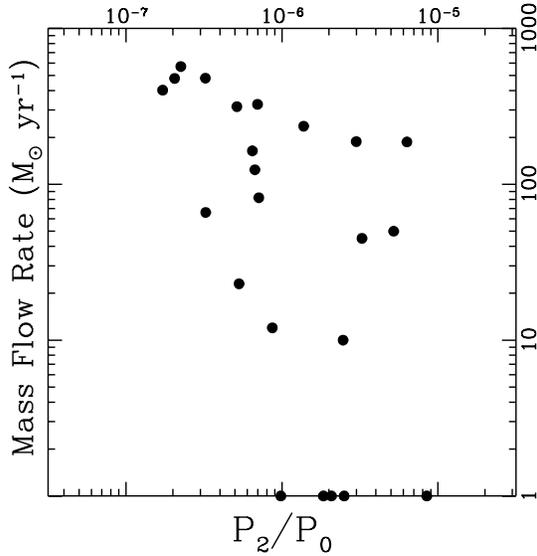

Fig. 8.—
Bautz-Morgan Type (see Table 1) vs $P_2/P_0$ computed in the $1h_{80}^{-1}$ Mpc aperture for all clusters in our sample (left) and those corresponding to the Edge et al. (1990) sample (right).

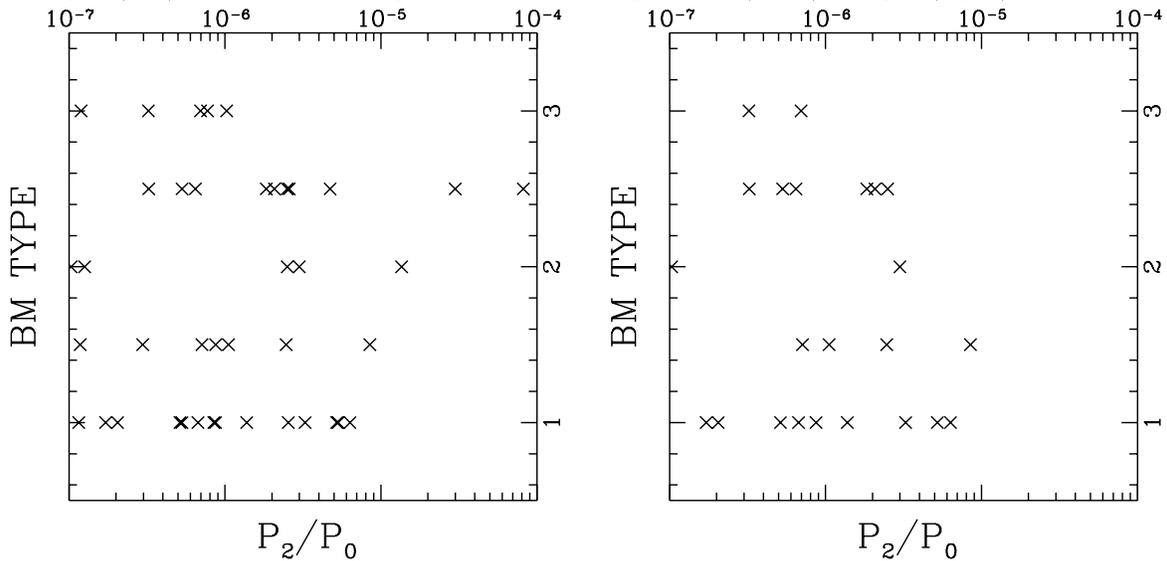